\documentclass[a4paper]{memoir}\setsecnumdepth{subsection}\settocdepth{subsection}
\makeheadrule{headings}{\textwidth}{0.3pt}
\copypagestyle{fnsizeheadings}{headings}
\makeevenhead{fnsizeheadings}{\thepage}{}{\footnotesize\slshape\leftmark}
\makeoddhead{fnsizeheadings}{\footnotesize\slshape\rightmark}{}{\thepage}

\usepackage[utf8]{inputenc}
\usepackage[american,ngerman,greek,brazilian,british]{babel}

\usepackage[sc,osf]{mathpazo}
\usepackage[scaled=0.875]{helvet} 
\usepackage[T1]{fontenc}
\usepackage{lmodern}
\usepackage[scaled=0.86]{berasans}
\usepackage[scaled=1.03]{inconsolata}
\usepackage{setspace}

\usepackage{graphicx}
\usepackage[babel=true]{microtype}
\usepackage{amsmath,amssymb,amsthm}
\usepackage{mathtools}
\usepackage{icomma}
\usepackage{xcolor}
\definecolor{darkGreen}{RGB}{0,110,0}
\definecolor{darkBlue}{RGB}{0,0,130}
\usepackage[colorlinks,citecolor=darkGreen,linkcolor=darkBlue,%
urlcolor=blue,hyperindex]{hyperref}
\usepackage{bm}
\usepackage{verbatim}
\usepackage[multiple,stable]{footmisc}
\usepackage{footnote}	
\makesavenoteenv{minipage}
\usepackage{csquotes}
\usepackage{booktabs}
\usepackage{multirow}
\usepackage[mathscr]{euscript}
\usepackage{bookmark}
\usepackage{mdwlist}
\usepackage{pdfpages}
\usepackage{subcaption}

\usepackage[normalem]{ulem}
\useunder{\uline}{\ul}{}

\usepackage{float}

\usepackage{comment}

\usepackage{tikz}

\usepackage{etoolbox}

\makeatletter
\pretocmd{\@schapter}{\setcounter{footnote}{0}}{}{}
\pretocmd{\@chapter}{\setcounter{footnote}{0}}{}{}
\makeatother

\DeclareMathOperator{\tr}{Tr}

\let\originalleft\left
\let\originalright\right
\renewcommand{\left}{\mathopen{}\mathclose\bgroup\originalleft}
\renewcommand{\right}{\aftergroup\egroup\originalright}

\newcommand{\exval}[3]{\left\langle #1 \middle| #2 \middle| #3 \right\rangle}

\newcommand{\bnorm}[1]{\bigl\|#1\bigr\|}

\newcommand{\norm}[1]{\left\Vert#1\right\Vert}
\newcommand{\abs}[1]{\left\vert#1\right\vert}

\newcommand{\trace}[1]{\mbox{Tr}\left( #1 \right)}

\newcommand{\ket}[1]{\left\vert#1\right\rangle}
\newcommand{\bra}[1]{\left\langle#1\right\vert}
\newcommand{\braket}[2]{\left\langle#1\vert#2\right\rangle}
\newcommand{\ketbra}[2]{\left\vert #1 \,\rangle \! \langle #2 \right\vert}
\newcommand{\avg}[1]{\left\langle#1\right\rangle}

\newcommand{\eg}{\emph{e.g.}}
\newcommand{\ie}{\emph{i.e.}}

\newtheorem{teo}{Theorem}[section]

\newtheorem{defi}[teo]{Definition}

\makeatletter
\newtheorem*{rep@theorem}{\rep@title}
\newcommand{\newreptheorem}[2]{%
\newenvironment{rep#1}[1]{%
 \def\rep@title{#2 \ref*{##1}}%
 \begin{rep@theorem}}%
 {\end{rep@theorem}}}
\makeatother

\newreptheorem{teo}{Teorema}
\newreptheorem{lema}{Lema}

\definecolor{darkgreen}{rgb}{0.36,0.72,0.1}
\mathtoolsset{centercolon}

\title{Tomography on Continuous Variable Quantum States}
\author{Ludmila Augusta Soares Botelho}
\date{\today}
\chapterstyle{veelo}
\begin{document}
\begin{center}

\thispagestyle{empty}

\vspace*{10.5cm}
{\huge Tomography on Continuous Variable Quantum States}

\vspace{5.5cm}

{\LARGE Ludmila Augusta Soares Botelho}

\vspace{0.5cm}
\begin{Large}
August 2018
\end{Large}

\end{center}
\newpage

\thispagestyle{empty}

\begin{center}

\settowidth{\unitlength}{\LARGE Tomography on Continuous Variable Quantum States} 
\vspace*{\baselineskip}

\rule{\textwidth}{1.6pt}\vspace*{-\baselineskip}\vspace*{3.6pt}
\rule{\textwidth}{0.4pt}\\[\baselineskip]
{\Huge Tomography on Continuous Variable Quantum States\\ [0.05\baselineskip]}
\rule{\textwidth}{1.6pt}\vspace*{-\baselineskip}\vspace*{3.6pt}
\rule{\textwidth}{0.4pt}\\[\baselineskip]

\vspace*{1.5cm}
{\LARGE Ludmila Augusta Soares Botelho}\\[\baselineskip]
\vspace*{2cm}
{\Large {Orientador:}\\
\vspace*{0.5cm}
Prof. Dr. Reinaldo Oliveira Vianna}\par \vfill

\begin{minipage}{7.8cm}
\small{\textbf{Versão Final} - Disserta\c{c}\~{a}o apresentada à UNIVERSIDADE FEDERAL DE MINAS GERAIS - UFMG,
como requisito parcial para a obtenção do grau de MESTRE EM
F\'{I}SICA.}
\end{minipage}\\
\vspace*{1cm}
Belo Horizonte\\
Brasil\\
Agosto de 2018
\end{center}

\chapter*{Dedicate}

\thispagestyle{empty}

\bigskip\bigskip\bigskip\bigskip\bigskip\bigskip\bigskip\bigskip\bigskip\bigskip\bigskip\bigskip\bigskip\bigskip\bigskip\bigskip\bigskip\bigskip\bigskip\bigskip\bigskip\bigskip\bigskip\bigskip\bigskip\bigskip                                              
\begin{center}
To my mother Noely Evangelina Augusta de Oliveira (\emph{in memoriam}).\\
\end{center}

\thispagestyle{empty}

\newpage

\thispagestyle{empty}

\bigskip\bigskip\bigskip\bigskip\bigskip\bigskip\bigskip\bigskip\bigskip\bigskip\bigskip\bigskip\bigskip\bigskip\bigskip\bigskip\bigskip\bigskip\bigskip\bigskip\bigskip\bigskip\bigskip\bigskip\bigskip\bigskip                                              

\vspace*{\fill}
\epigraphfontsize{\itshape}
\epigraph{"I am among those who think that science has great beauty."}{--- \textup{Marie Skłodowska Curie }}

\chapter*{Acknowledgements}
\thispagestyle{empty}

Agradeço a minha mãe, dona Noely, para quem eu dedico está dissertação em homenagem a sua memória. Eu nada seria sem o exemplo de força e ímpeto dessa mulher incrível que inspirou a muitas pessoas. Ela sempre acreditou em meu potencial e investiu em mim e sem a mesma esse trabalho nem teria começado. Agradeço também ao meu irmão Abner, que é a minha família, estaremos sempre juntos. Agradeço também a Tia Graça, Tio Dinha, Tio Toca e Lélia, que me ajudaram muito.

Ao meu orientador Reinaldo O. Vianna, por ter me dado a oportunidade de trabalhar com um problema legal, pela paciência, e por ter descriptografado meus textos. Obrigada pelos ensinamentos, cujo fruto é esta dissertação.

Ao professor Mario Mazzoni, que desde a graduação faz qualquer assunto complicado entrar na cabeça de qualquer um. Agradeço também aos professores Carlos Henrique Monken, Jafferson Kamphorst e José Rachid Mohallem pelos ensinamentos. 

Aos meu amigão Marcello\footnote{Metaldumal}, que tem me dado muito apoio nos momentos mais tenebrosos da minha vida, muitos cigarros\footnote{Ainda bem que paramos de fumar}, alguns cafés. Sua presença centrada me acalma\footnote{Até na "estrada da morte" na "serra da crueldade".}. Você é muito importante para mim. Agradeço também ao meus veteranos Davi, Jéssica, Alana, Tati e Cobra, pelas festinhas, pelos cafés e boas conversas.

Á dupla Marco Túlio e Mateus Araújo, que foram os primeiros a falar comigo quando cheguei na física. Eles me mostraram o tanto que quântica pode ser divertida, tomo eles como exemplo de pesquisadores. Vocês foram e ainda são umas das principais influencias nesse processo todo.

Ao rapazes da república mais de boas que eu conheço: Balde, pelas viagens caleidoscópicas, Gil, por me incentivar na física e ótimas discussão sobre ensino e consciência de classe e raça, Olímpio, pelas conversas e ideias sensacionais e Marião, também pelas conversas, companhia, templates no \LaTeX, bebedeiras e super puxões de orelha\footnote{Muito merecidos.}.

Devo reservar um parágrafo para Enilse Esperança, que de repente apareceu em minha vida\footnote{E me "Nocauteou, me tonteou/Veio à tona, fui à lona, foi K.O."} e me traz grande alegria. Você cuida de mim, abre minha cabeça e me inspira todos os dias. <3

Ao pessoal do dojo Yamashi Castelo, Sensei Davi, Lucas, Tiagão e Tatá. Treinar com vocês é uma das coisas que me ajudou a não pirar, além de me deixar mais forte\footnote{Físico e psicologicamente}. E a galera do Ubuntu Rugby Club\footnote{"Sou o que sou pelo que nós somos!"}, em especial ao Didi, Nati, Kelly, Dalton, Davi, Daisy e Daiane, pelos bons treinos e companheirismo.

A minha amiga Bárbara Diniz, pelos rolês, rangos, conversas e ponderações. A senhora ajudou mais do que você imagina em um monte de coisas.  

Agradeço a minha xará, Ludimila Franciane, pelo companheirismo, pela paciência, e por ter me ajudado em um momento muito difícil.

À galera da salinha do mestrado: Clóvis, Bel, Jéssica, Saulo, Geovani, Tiago, Tamires, Rafael, João e Monalisa. Estudar com vocês me ajudou muitíssimo pra aprender melhor além de ter deixado minhas tardes e cafés mais agradáveis.

Um agradecimento especial aos rapazes do Infoquant\footnote{Cafofeiros}: Diego, por ter me ensinado a montar meus primeiros algoritmos, ao Thiago "Tchê"\footnote{É tão "Dungeon Master", que "mestra" até o pessoal no cafofo}, que me ensinou e ensina todos nós, não seriamos nada sem você, ao Lucas pelas boas indicações de leituras e pelo seu vasto conhecimento, não somente de física, mas sempre com as melhores curiosidades da vida, ao Tanus pela enorme sabedoria e dicas de problemas, ao Léo, ao Felipe e ao João. Me sinto muito feliz e acolhida no nosso ambiente de trabalho, eu aprendo muito e dou boas risadas com vocês. Os cafofeiros seniores, Debarba e Iemini, agradeço pelas excelentes discussões nas poucas vezes que nos vimos. Também devo agradecer a galera do Enlight, Denise, Davi, Sheila, Marina, João e Raul, pelos cafés e zoeiras na salinha.

Não posso deixar de agradecer aos meus amigos das antigas, como o Lucas Humberto, que sempre foi um grande companheiro irmão e confidente. À Angela e Humberto, me sinto praticamente filha de vocês. A galera do teatro: Tarcísio, Priscila, Zilah e Marcela (in memorian), que fez com que 2008\footnote{O período da inocência} nunca acabasse, vocês são eternos, viva nossos 10 anos de amizade! Agradeço também ao Pep e a Lucimara, as meninas da Terra de Godart, pelas loucuras, aleatoriedades e bebedeiras.\vfill
\noindent -- The author does not consider this work as completed due to problems after its presentation and lack of review. Suggestions for improvement and error notes can be sent to ludmilaasb@gmail.com\\
-- This text was written by a student for students.\\
-- Este trabalho teve o apoio financeiro direto e indireto da \textbf{CAPES}, da \textbf{FAPEMIG} e do \textbf{CNPq}.

\chapter*{Abstract}

In this work we have explored few tools in Quantum State Tomography for Continuous Variable Systems. The concept of quantum states in phase space representation is introduced in a simple manner by using a few statistical concepts. Unlike most texts of Quantum information in which the Wigner function for a single mode is often more used, in this text the multi-modes state Wigner function is also developed. Our numerical investigations indicate that the reconstructed method using back-projection add some error due the choice of cutoff frequency, therefore it is necessary to use data post-processing, like the semi-definite programs, which provides sufficient conditions correctly estimate the state. Once the information about the state is recovered, important features such as entanglement can also be investigated. \newline\newline\newline\newline\newline\newline\newline\newline\newline\newline

\subsubsection*{Keywords}

\textit{Wigner function, quadrature, continuous variable, Gaussian state, coherent state, squeezed state, Fock state, single mode state, multi mode state, homodyne detection, tomography, Radon transform, inverse Radon, back-projection algorithm, kernel, cutoff, characteristic function, fidelity, semi-definite programs, SDP, entanglement.}

\chapter*{Resumo}

Neste trabalho exploramos algumas ferramentas da Tomografia de Estados Quânticos em sistemas de Variáveis Contínuas. O conceito de estados quânticos na representação do espaço de fase é introduzido em uma simples abordagem utilizando um pequeno número de conceitos estatísticos. Ao contrário da maioria dos textos em Informação Quântica no qual a função de Wigner de estado de um modo é mais usual, neste texto a função de Wigner multi-modos é explorada. Nossa investigação numérica aponta o método de reconstrução utilizando o algoritmo de back-projection adiciona erro devido a escolha da frequência de corte, sendo assim é necessário utilizar pós processamento dos dados, como programas semi definidos, que provem condições suficientes para estimar corretamente o estado. Uma vez que a informação sobre o estado é recuperada, características importantes como o emaranhamento também podem ser investigadas.\newline\newline\newline\newline\newline\newline\newline\newline\newline\newline

\subsubsection*{Palavras-chave}

\textit{Função de Wigner, quadratura, espaço de fase, variáveis contínuas, estado gaussiano, estado coerente, estado squeezed, estado de Fock, estados de um mode, estados multimodo, detecção homódina, tomografia, transformada de Radon, Radon inversa, algoritmo de back-projection, kernel, cutoff, função característica, fidelidade, programas semidefinidos, SDP, emaranhamento.}

%
\clearpage
\tableofcontents
\clearpage
\newpage
\listoffigures                 
\newpage
%
\chapter*{Introduction}



\epigraphfontsize{\small\itshape}
\epigraph{``DON'T PANIC!''}{--- \textup{Douglas Adams} \\ The Hitchhiker's Guide to the Galaxy}

How to write something that is infinite? And how to reconstruct it?

It seems that's a very difficult task, since we need infinite "things" to compute. But don't lose your hope! Answering the first question, thanks to very smart people, we can write in a piece of paper something that symbolize those infinite ``things'' in short lines. Let's talk about continuous variables. They can take on infinitely many, uncountable values, \ie, we can't even order it. But, who said it needs to put them in a explicit form?

Through the graduation on Physics, we get familiar with continuous variables and continuous functions: from calculus classes we learn about the set of real numbers, for example. Think about all the numbers between zero and one. How should we write then? We can't, they are uncountable, unlike the natural numbers. Or draw a line on a paper sheet without taking the pencil away: it can be a representation for a continuous function. \footnote{However, we know the pencil is just spreading graphite, made of atoms of carbon, which means in reality it IS discretized. Still, it is a good approximation for our senses.}

The functions of continuous variables are present all the time on physics and mathematics, therefore we have special tools to deal with them like limits, derivatives, etc. Moreover, we use it to describe states on classical mechanics and probabilistic distribution on statistics. 

Besides the very elegant Dirac's representation and the usefulness of linear algebra, the ``old'' quantum mechanics was based on continuous variables functions, if we think about the concept of ``wave function'', for example.

After while, we had tons of research on discrete, low dimension quantum systems. Their matrices are easy to write by hand and to check some proprieties also. For example, if we think about entanglement, it becomes harder very fast if one increases the system partitions and/or dimensions.
 
Moreover, continuous variable systems are very useful: they have this ``robustness'', they are feasible on laboratory, such as \textit{Gaussian States} on quantum optics. 

Although, we still have the second question to answer. That's a little bit trickier, if you want a full reconstruction of the state, you need to perform on every bases elements. Since it is impossible measuring infinite things, our information is aways incomplete! But, maybe you don't need to measure all the infinite to get the information you want. 

I like to think on photography: take a picture of the Da Vinci famous painting Mona Lisa. Digital cameras codifies the information of this ``continuous function'' on pixels, which are discretized. If the camera is good enough, we have the feeling of a very reliable representation. Although, if you zoom it, you can see the colorful, tiny, different squares. It's about resolution and what information do you want. Maybe, even with low resolution, you can say it is Mona Lisa and not the Johannes Vermeer's Girl With a Pearl Earring. \footnote{The paints have completely different styles, besides been woman portrait, it's quite obvious the difference.}    

\begin{figure}
\centering
 \includegraphics[width=0.5\textwidth]{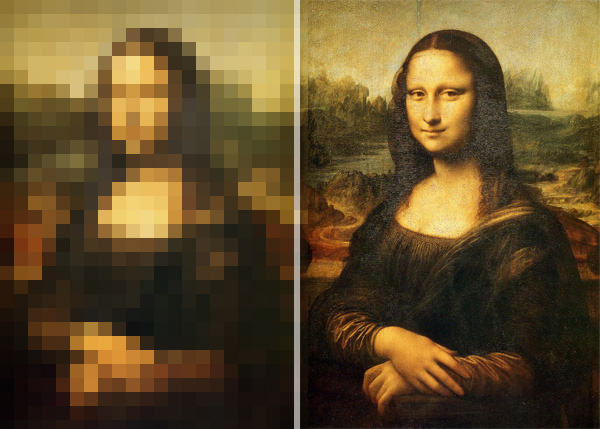}
 \caption{How do you know it is Mona Lisa?}
\end{figure}

On this dissertation, I want to give to the reader a simple approach to how to deal with continuous variable systems and a toolbox for tomographic reconstruction of a state. Moreover, I will discuss a little on entanglement and the efficiency of reconstructions algorithms. I hope you enjoy! 
\chapter{Writing the Infinite: Dealing with Continuous Variables}



\epigraphfontsize{\small\itshape}
\epigraph{"Mathematics is a game played according to certain rules with meaningless marks on paper."}{--- \textup{David Hilbert}}

A quantum state is usually described by its \textit{density matrix} (or density operator) $\rho$. Such object lives in a complex Hilbert space $\mathcal{H}$ and needs to satisfy the following conditions \cite{ballentine1998quantum}:

\begin{align}
(i)\; &\text{\textit{Hermitian,}} & &\rho = \rho^\dagger; \\
(ii)\; &\text{\textit{positive semi-definte,}} & &\rho \geq 0;\\
(iii)\; &\text{\textit{normalized}} & &\trace{\rho} = \bnorm{\rho}_1 = 1 .
\end{align}

There is a special class of states, the \textit{pure states}, $\rho=\ketbra{\Psi}{\Psi}$, where the unit-norm state $\ket{\Psi}$ is named \textit{state vector}.

Considering a continuous variable scenario, the operators associated with the system degrees of freedom have continuous spectrum. Since the bases of the those operators eigenstates form an infinite-dimensional Hilbert space $\mathcal{H}$ of the system, the \emph{explicit} matrix elements representation of $\rho$ is not possible. However, you can still write it in a piece of paper, in fact, in a similar way we already do with states in classic mechanics.

The usual representations are the position and the momentum. There is also the quadrature representation, which combines position and momentum and is quite useful to study, \eg, electromagnetic field modes. In this chapter, we are going to talk about an useful tool, the Wigner Function, introduced by Wigner on his original article \footnote{The original Wigner article is about quantum corrections to classical statistical mechanics where Boltzmann factors contain the energies which in turn are expressed as functions of both $q$ and $p$ } from 1932 \cite{Wigner1932}. It provides an equivalent representation of any quantum state in the quadrature phase space, in a sense to retrieve the idea of probability distribution. Since it accept some negativeness,  it is not really a probability density function, but has similar proprieties, works similarly to a weight function. We start reminding some ideas of statistical concepts and then we derive the function, illustrating with special examples of continuous variable states, the Gaussian States, and the relation with tomography.

\section{Wigner Function}

If $X$ is a random variable, we define the characteristic function $\Phi_X(t)$ as the mean value of $e^{itX}$, with $t$ as a real number:

\begin{equation}
\Phi_X (t) = \langle e^{(itX)} \rangle; \qquad	t \in \mathbb{R}.
\end{equation}

Given a characteristic function, we can build a probability density function:

\begin{equation}
F_t(x) = \frac{1}{2\pi} \int e^{-itx} \Phi_X(t) \mathrm{d}t.
\end{equation}

Now, let's try to build a probability distribution associated to the phase space, for quantum operators. Instead of a random variable now we have the pair $(q,p)$ - position and momentum. The characteristic function associated to that pair of random variables would be:
\begin{equation}
\avg{e^{i(t_1q + t_2p)}}
\end{equation}
where $t_1$ and $t_2$ are real. Let $t_1=-u$ and $t_2=-v$, with $u,v \in \mathbb{R}$ \footnote{This choice is a convenience}. It's important to note that changing $q$ and $p$ for the operators $\hat{q}$ and $\hat{p}$, \(e^{-i(u\hat{q}+v\hat{p})/\hbar}\) is an operator that makes a translation on phase space, and it's called \textit{Weyl Operator}.

%
%
%

The expected value for the Weyl Operator, given a state $\rho$, defines the characteristic function:
\begin{equation} \label{eq:charfun}
\widetilde{W}(u,v) = \tr[\rho e^{-i(u\hat{q} + v\hat{p})/\hbar}],
\end{equation}
and the associated \textit{probability density function}:
\begin{equation} \label{eq:probabilitydensfun}
W(q,p) =  \left(\frac{1}{2\pi\hbar}\right)^2 \iint  \widetilde{W}(u,v) e^{i(uq + vp)/\hbar}\mathrm{d}u\mathrm{d}v.
\end{equation}

This is the \textit{Wigner Function}, a Fourier transform of the characteristic function. On the other hand, given a Wigner Function, one can invert the Fourier transform and find the characteristic function as well.

We want to write the Wigner function in a more explicit form. To do so, we need to work with the characteristic function in a more convenient way.

Given the operators $A$ and $B$, such that $[A,[A,B]] = [B,[A,B]] = 0$, we recall the Baker–Hausdorff formula:
\begin{equation}
e^{A+B} = e^A e^B e^{-[A,B]}. \\ \label{eq:bakerhaussdorf}
\end{equation}
Using the commutation relation of the operators $\hat{q}$ and $\hat{p}$, $[\hat{q},\hat{p}]= i\hbar$, we have
\begin{equation}
e^{(-u\hat{q}-iv\hat{p})/\hbar} = e^{-iu\hat{q}/\hbar} e^{-iv\hat{p}/\hbar} e^{iuv/2\hbar}.
\end{equation}

From this, we use the identity $\mathbb{1} =\int\ket{q}\bra{q} \mathrm{d}q$:

\begin{equation}
\begin{split}
e^{iuv/{2\hbar}} \int e^{-iu\hat{q}/\hbar} e^{-iv\hat{p}/\hbar} \ket{q}\bra{q} \mathrm{d}q &= e^{iuv/{2\hbar}} \int e^{-iu\hat{q}/\hbar} \ket{q+v}\bra{q} \mathrm{d}q  \\ 
 &= e^{iuv/{2\hbar}} \int e^{-iu(q+v)/\hbar} \ket{q+v}\bra{q} \mathrm{d}q .
\end{split}
\end{equation}

Let $q+v = q'+\frac{v}{2}$, therefore $\mathrm{d}q = \mathrm{d}q'$ and $q=q'-\frac{v}{2}$. We have then:
\begin{equation}
\begin{split}
e^{-iu\hat{q}-iv{p}/\hbar} &= e^{iuv/{2\hbar}}\int e^{-iu(q'+\frac{v}{2}/\hbar)} \ket{q+\frac{v}{2}} \bra{q'-\frac{v}{2}} \mathrm{d}q' \\
 &=\int e^{-iuq'/\hbar} \ket{q+\frac{v}{2}} \bra{q'-\frac{v}{2}} \mathrm{d}q' \label{eq:truque1}.
\end{split} 
\end{equation}

Equation (\ref{eq:charfun}) can be rewritten using (\ref{eq:truque1}) as:
\begin{align}
\widetilde{W}(u,v)&= \iint \mathrm{d}q \mathrm{d}q' \bra{q} \rho e^{-iuq'/\hbar} \ket{q+\frac{v}{2}} \bra{q'-\frac{v}{2}} \ket{q} \nonumber \\
&=\iint \mathrm{d}q \mathrm{d}q' \bra{q} \rho e^{-iuq'/\hbar} \ket{q+\frac{v}{2}} \delta \left[ \left( q'-\frac{v}{2}\right)-q \right] \nonumber \\
&=\int\mathrm{d}q' \bra{q'-\frac{v}{2}} \rho \ket{q'+\frac{v}{2}} e^{-iuq'/\hbar}.
\end{align}

Therefore, the probability density function (\ref{eq:probabilitydensfun}) is: 
\begin{equation}
W(q,p) = {\left(\frac{1}{2\pi\hbar}\right)}^2 \iint \mathrm{d}u \mathrm{d}v \int\mathrm{d}q' \bra{q'-\frac{v}{2}} \rho \ket{q'+\frac{v}{2}} e^{iu(q-q')/\hbar}e^{ivp/\hbar}.
\end{equation}

From the \textit{Dirac's Delta} definition:
\begin{equation} \label{eq:deldirac}
\frac{1}{2\pi\hbar} \int e^{iu(q-q')/\hbar} \mathrm{d}u = \delta(q-q'),
\end{equation}

we have: 
\begin{equation}
\begin{split}
W(q,p) =& \frac{1}{2\pi\hbar} \iint \bra{q'-\frac{v}{2}}\rho\ket{q'+\frac{v}{2}}\delta(q-q')e^{ivp/\hbar} \mathrm{d}v \mathrm{d}q'\\
=& \frac{1}{2\pi\hbar} \int \bra{q-\frac{v}{2}}\rho\ket{q+\frac{v}{2}} e^{ivp/\hbar} \mathrm{d}v .\label{eq:wignerfunction}
\end{split}
\end{equation}

The Wigner function can also be obtained using the momentum representation:
\begin{equation}
W(q,p) = \frac{1}{2\pi\hbar} \int \bra{p - \frac{u}{2}} \rho \ket{p+\frac{u}{2}} e^{iuq/\hbar} \mathrm{d}u.
\end{equation}

One of the Wigner function advantages, besides the graphic representation, is its marginal distributions yield the usual position and momentum probability distributions

\begin{align}
\int W(q,p) \mathrm{d}p &= \bra{q} \rho \ket{q}, \\
\int W(q,p) \mathrm{d}p &= \bra{p} \rho \ket{p}.
\end{align}

Let us check, for example, the position marginal:
\begin{equation*}
\int W(q,p) \mathrm{d}p = {\left(\frac{1}{2\pi\hbar}\right)}^2 \iint \bra{q-\frac{v}{2}} \rho \ket{q+\frac{v}{2}} e^{ivp/\hbar} \mathrm{d}v \mathrm{d}p.
\end{equation*}
Using again (\ref{eq:deldirac}):
\begin{align*}
\int W(q,p) \mathrm{d}p &= \int \bra{q-\frac{v}{2}} \rho \ket{q+\frac{v}{2}} \delta(v) \mathrm{d}v  \\
&= \bra{q} \rho \ket{q} .
\end{align*}

It is easy to see that $W(q,p)$ is correctly normalized
\begin{align*}
\iint W(q,p) \mathrm{d}q \mathrm{d}p &= \int \bra{q} \rho \ket{q} \\
&= \mathrm{Tr}(p) =1.
\end{align*}

Now, for analogy, we defined the Wigner function of an arbitrary operator\footnote{Notice that we don't have the factor $\frac{1}{2\pi\hbar}$ at the definition of $W_R$, it's just to simplify the notation} $R$:

\begin{equation}\label{eq:qualquerop}
W_R (q,p) = \int_{-\infty}^{+\infty} \bra{q-\frac{v}{2}} R \ket{q+\frac{v}{2}} e^{ipv/\hbar} \mathrm{d}v.
\end{equation}

The averaged value of an operator in Wigner's representation \footnote{This expression has the form of mean value on classic phase space} is:
\begin{equation}\label{eq:overlap}
\avg{R} = \mathrm{Tr}(\rho R) = \iint W(q,p) W_R(q,p) \mathrm{d}q \mathrm{d}p
\end{equation}

Note that in the classical case, $W(q,p)$ would be a probability density function ( $W(q,p)\geq 0$). Since $W(q,p)$ can assume negative values, we call it a \textit{quasi-probability}.

Changing the expression above using $W_R = 2\pi \hbar W_{\rho'}$,

\begin{equation}
\mathrm{Tr}(\rho \rho') = 2 \pi \hbar \iint  W_\rho W_{\rho'} \, \mathrm{d}q \mathrm{d}p.
\end{equation}
For any state operator $\rho$ and $\rho'$, we have \cite{ballentine1998quantum}
\begin{equation}
0 \leq \tr(\rho \rho') \leq 1.
\end{equation}

It means:
\begin{equation}
0 \leq \iint W_\rho W_\rho' \,\mathrm{d}q \mathrm{d}p \leq \frac{1}{2\pi \hbar},
\end{equation}
with the upper limit reached if and only if $\rho = \rho'$ is a pure state operator. This is specially useful because it allows us to quantify the \textit{purity} of quantum state, $\mathrm{Tr}(\rho^2)$.

Finally, we can use the relation (\ref{eq:overlap}) to represent the density-matrix elements using elements in a given basis in terms of the Wigner function

\begin{equation}\label{eq:matrixind}
\exval{a'}{\rho}{a} = \mathrm{Tr}(\rho \ketbra{a}{a'}) = 2 \pi \hbar \iint  W_\rho W_{a'a} \, \mathrm{d}q \mathrm{d}p,
\end{equation}
with $W_{a'a}$ being the Wigner representation of the projector $\ketbra{a}{a'}$, and it is obtained changing $R$ to the projector in eq. (\ref{eq:qualquerop}).

There is another way of making quantum-mechanical predictions, that is, of calculating expectation values via Wigner functions. We can associate it with the moments of the characteristic function through this relation \cite{bengtsson2017geometry}:

\begin{equation}
\tr \rho (u\hat{q} + v\hat{p})^k = i^k \left( \frac{\mathrm{d}}{\mathrm{d} \sigma} \right)^k \tr \rho e^{i\sigma(u\hat{q} + v\hat{p})} \mid_{\sigma=0} =i^k  \left (\frac{\mathrm{d}}{\mathrm{d} \sigma} \right)^k \widetilde{W}( \sigma u, \sigma v)\mid_{\sigma=0}
\end{equation}
But if we undo the Fourier transformation we have
\begin{equation}
\tr \rho (u\hat{q} + v\hat{p})^k = \frac{1}{2 \pi \hbar} \int \mathrm{d}q \mathrm{d}p (uq + vp)^k W(q,p).
\end{equation}

By comparing the coefficients we see that the moments of the Wigner function give the expectation values of symmetrized products of operators, that is to say:

\begin{equation}
\tr \rho (\hat{q}^m \hat{p}^n)_{sym}, 
\end{equation}

where $(\hat{q}\hat{p})_{sym}$ means that we should symmetrize all possible products of the $m$ $\hat{q}$-operators and the $n$ $\hat{p}$-operators. 

\subsection{Wigner Multipartite}

Let us consider a system with $n$ canonical degrees of freedom. It could be $n$ harmonic oscillators or $n$ electromagnetic field modes. The canonical commutation relations between $2n$ self-adjoint operators of such system could be easily described using the vector:

\begin{equation}
\hat{O} = (\hat{O}_1,...,\hat{O}_{2n})^T = (\hat{q}_1,\hat{p}_1,...,\hat{q}_n,\hat{p}_n)^T . \label{eq:vectoroperator}
\end{equation}	

%

With this parametrization, the commutation relations have the form:
\begin{equation}
[\hat{O}_j,\hat{O}_k] = i\hbar \sigma_{jk},
\end{equation}
being the $\sigma$ $2n \times 2n$, symmetric and bloc diagonal, called symplectic matrix, defined by:

\begin{equation}\label{eq:sigma}
\sigma = \bigoplus_{j=1}^n \begin{bmatrix} 
0 & 1\\
-1 & 0
\end{bmatrix}.
\end{equation}
The phase space then is equipped with a symplectic form and it is isomorphic to $\mathbb{R}^{2n}$. We want to expand our Wigner representation for a composite system. To do so, let us define the Weyl operator, now for a multi-mode system:
\begin{equation}
\mathcal{W}_\xi = e^{-i \xi^T \hat{O}},
\end{equation}
for $ \xi \in \mathbb{R}^{2n}$, our characteristic function is then:

\begin{equation}
\widetilde{W}(\xi) = \mathrm{Tr}[\rho \mathcal{W}_\xi].
\end{equation} 

Each characteristic function is uniquely associated with a state through a Fourier-Weyl transform. One can show that the state $\rho$ is directly obtained from:

\begin{equation}
\rho = \frac{1}{(2\pi\hbar)^{2n}} \int \widetilde{W}(\sigma \xi) \mathcal{W}(-\sigma \xi) \mathrm{d}^{2n} \xi .
\end{equation}

For simplicity, let us consider the case for two mode state. The result can be easily extended to more modes. The two mode Weyl operator is\footnote{The operators labels makes implicit were they act non-trivially, \eg , $\hat{q}_1 = \hat{q} \otimes \mathbb{1}$.}: 
\begin{equation}
\mathcal{W}_\xi = e^{-i[(u_1\hat{q}_1 + v_1\hat{p}_1) + (u_2\hat{q}_2 + v_2\hat{p}_2)]/\hbar}, \label{eq:bipartweyl}
\end{equation}
for $\xi = (u_1,v_1,u_2,v_2)^T$. Note that we can separate the operator above using the Baker-Haussdorf formula (\ref{eq:bakerhaussdorf}) two times and the commutations relation given by (\ref{eq:sigma}):

\begin{equation*}
 \left( e^{-i(u_1\hat{q}_1 +v_1\hat{p}_1 )/\hbar}\right) \times \left(e^{-i(u_2\hat{q}_2 +v_2\hat{p}_2)/\hbar} \right).
\end{equation*}

From here, we compute the equation above the same way done before for a single mode. The completeness relation for the Hiblert space of two modes $\mathcal{H}_1 \otimes \mathcal{H}_2$ is\footnote{Here, we shortened the notation for $\ket{q_1, q_2} = \ket{q_1}\otimes \ket{q_2}$} 

\begin{equation*}
\mathbb{1}_{\mathcal{H}_1 \otimes \mathcal{H}_2} = \int \ket{q_1, q_2}\bra{q_1, q_2} \mathrm{d}q_1 \mathrm{d}q_2 .
\end{equation*}

Since the operators $O_n$ act on  the corresponding labeled space, we can compute our two-mode characteristic function:
\begin{equation}
\begin{split}
\widetilde{W}(u_1,v_1,u_2,v_2) = & \iint e^{-i(u_1q_1'+ iu_1q_1')/\hbar} \\
&\times \bra{q_1-\frac{v_1}{2},q_2-\frac{v_2}{2}} \rho \ket{q_1+\frac{v_1}{2},q_2+\frac{v_2}{2}} \\
&\times \mathrm{d}q_1 \mathrm{d}q_2
\end{split}
\end{equation}
and bipartite Wigner representation:
\begin{equation}
\begin{split}
W(q_i,p_i,q_j,p_j) = & \frac{1}{(2\pi\hbar)^2} \iint e^{i(v_1p_1 + v_2p_2)/\hbar} \\
& \times \bra{q_1-\frac{v_1}{2},q_2-\frac{v_2}{2}} \rho \ket{q_1+\frac{v_1}{2},q_2+\frac{v_2}{2}} \\
& \times \mathrm{d}v_1 \mathrm{d}v_2 
\end{split}
\end{equation}

\section{Gaussian States}

Gaussian functions are widely use throughout the study of probability and statistics, often called``normal distributions''. Usually, on a physics or mathematics course they are naturally introduced, since it has various applications and utility. In a phase space descriptions, Gaussian states are characterized through their property that the characteristic function is a Gaussian. They are efficiently producible in the laboratory, \eg, \textit{coherent state}, such as those from a laser, thermal states and vacuum states.

From the previous section formalism for a quantum system with $n$ canonical degrees of freedom, our multi-mode characteristic Gaussian is \cite{Eisert2003,Wang2008}: 
\begin{equation}
\widetilde{W}_\rho(\xi) = \widetilde{W}_\rho(0)e^{-\frac{1}{4} \xi^T \Gamma \xi + D^T \xi},  
\end{equation}
where is $\Gamma$ a $2n \times 2n$-matrix and $D \in \mathbf{R}^{2n}$ is a vector. A Gaussian characteristic function can be characterized via its first and second moments alone as consequence, \ie, it is possible to describe such states in terms of finite-dimension matrices. It means that, a $n$ mode Gaussian state requires only $2n^2 + n$ real parameters for its full description \cite{Eisert2003}. The first moments form a vector $d \in \mathbb{R}^{2n}$, the displacement vector:
\begin{equation}
d_j = \tr\left[ O_j \rho \right],
\end{equation} 
where $j =1, \dots , 2n $. They are linked to the above $D$ by $D = \sigma d$, with $\sigma$ been the sympletic matrix (\ref{eq:sigma}), been the expected values of the canonical mode operators. The second moments, which form a real symmetric $2n \times 2n$ covariance matrix $\gamma$, are defined as:

\begin{equation}
\gamma_{j,k} = 2 \mathrm{Re} \{\tr \rho [O_j - Tr(O_j \rho)][O_k - Tr(O_k \rho)]  \}.
\end{equation}
The link with $\Gamma$ is $\Gamma = \sigma^T \gamma \sigma$. It is important to notice that a quantum state needs to respect a Heinsenberg uncertainty relations, hence not any real symmetric $2n \times 2n$-matrix can be a legitimate covariance of a quantum state. In terms of a covariance matrix, the uncertainty than can be written as: 

\begin{equation} \label{eq: incerteza}
\gamma i\sigma \geq 0 .
\end{equation}
%
%
%
In other words, we can say that for any real symmetric matrix $\gamma$ satisfying the equation (\ref{eq: incerteza}) a Gaussian state whose covariance matrix is $\gamma$ \cite{Eisert2003} exist.

\subsection{Coherent State}
In quantum optics the coherent state refers to a state of the quantized electromagnetic field, which has dynamics most closely resembling the oscillatory behavior of a classical harmonic oscillator. The state of a light beam out of a laser device is a coherent state \cite{Glauber1963}.

Since we are talking about light, let us remind about the Fock States, which are very useful for understand better the coherent states. Describing a quantum state through the ``number of photons\footnote{Which are identical and have bosonic nature}'' is to move the state address from the Hilbert Space to the Fock Space, for a more suitable representation, considering that describes an infinite vector space but now it is quantized and enumerable. After this short refresher, we will move back to phase space representation.

A Fock state, denoted by $\ket{n}$ is a eigenstate of the photon-number operator $\hat{n} = \hat{a}^{\dagger}\hat{a}$, where $n$ represents a fixed photon number.

The annihilation operator $\hat{a}$ and creation operator $\hat{a}^{\dagger}$, lowers or raises the photon number in integer steps
\begin{align}
\hat{a}\ket{n} &= \sqrt{n}\ket{n-1}, \\
\hat{a}^{\dagger}\ket{n} &= \sqrt{n+1}\ket{n+1},\label{eq:creatorop}
\end{align}
and for a state with zero photons, the annihilation operator acts $\hat{a}\ket{0} = 0$. We call the state $\ket{0}$ as \textit{vacuum state}. From it and the relation \ref{eq:creatorop} one can write an $\ket{n}$ like
\begin{equation}
\ket{n} = \frac{\hat{a}^{\dagger n}}{\sqrt{n!}} \ket{0}.
\end{equation}

The $\hat{q}$ and $\hat{p}$ operators can be expressed using the annihilation and creation operators\footnote{Those operators will be explained in the next section.}:

\begin{equation}
\hat{q} = (\hat{a} + \hat{a}^\dagger)/2, \qquad \hat{p} =-i(\hat{a} - \hat{a}^\dagger)/2. \label{eq:anihhcreation}
\end{equation}

And we can obtain the formula for their space representation, for a single mode:

\begin{equation}
\psi_n(q) = \frac{H_n(q)}{\sqrt{2^n n! \sqrt{\pi}}} \exp\left(\frac{-q^2}{2}\right), \label{fockhermite}
\end{equation} 
where $H_n$ denote the Hermite polynomials. Note that for the vacuum state we have:
\begin{equation}
\psi_0(q) = \pi^{-1/4} \exp\left(\frac{-q^2}{2}\right), \label{eq:psivacuum}
\end{equation}
and, as we can see, it is an obvious Gaussian state. Fock states form a complete set,
\begin{equation}
\sum_{n=0}^\infty \ket{n}\!\bra{n} = 1,
\end{equation}
and orthonormal because they are eigenstates of the Hermitian operator $\hat{n}$.

We can define the coherent states as the eigenstates of the annihilation operator $\hat{a}$
\begin{equation}
\hat{a}\ket{\alpha} = \alpha \ket{\alpha}. \label{eq:coherent}
\end{equation}
Note that a vacuum state is also a coherent state, since it satisfies (\ref{eq:coherent}) for $\alpha = 0$.  The coherent states, as eigenstates of the annihilator operator $\hat{a}$, have well-defined amplitudes, $\norm{\alpha}$, and phases, arg $\alpha$. Because $\hat{a}$ is not Hermitian, its eigenvalues are complex. 

Using the \textit{Fock state} representation, one can write the coherent state as: 
\begin{equation}
\ket{\alpha} = \exp\left(-\frac{1}{2}\abs{\alpha}^2 \right) \sum_{n=0}^{\infty} \frac{\alpha^n}{\sqrt{n!}} \ket{n}.\label{eq:coherentfock}
\end{equation}

To understand better those states, let us introduce the \textit{displacement operator} 
\begin{equation}
\hat{D}(\alpha) = \exp(\alpha \hat{a}^{\dagger} - \alpha^{*} \hat{a}), \label{eq:displamentop}
\end{equation}
which is unitary and displaces the amplitude $\hat{a}$ by the complex number $\alpha$
\begin{equation}
\hat{D}(\alpha)^{\dagger}\hat{a}\hat{D}(\alpha) = \hat{a} + \alpha. \label{eq:displaceaction}
\end{equation}

The proof of eq. (\ref{eq:displaceaction}) can be found on \cite{leonhardt2005measuring}. We can define a coherent state as a displaced vacuum:
\begin{equation}
\ket{\alpha} = \hat{D}(\alpha)\ket{0}.
\end{equation}

If we decompose the complex amplitude $\alpha$ into real and imaginary parts like
\begin{equation}
\alpha = 2^{-1/2}(q_0 + ip_0),
\end{equation}
represent the displacement operator in terms of $\hat{q}$ and $\hat{p}$,  
\begin{equation}
\hat{D} = \exp (ip_0\hat{q}-iq_0\hat{p}),
\end{equation}
and separating it using Baker-Hausdorff formula (\ref{eq:bakerhaussdorf}),
\begin{equation}
\begin{split}
\hat{D} = & \exp\left(-\frac{ip_0q_0}{2}\right)\exp\left(ip_0\hat{q}\right)\exp\left(-iq_0\hat{p}\right)\\
= & \exp\left(+\frac{ip_0q_0}{2}\right)\exp\left(-iq_0\hat{p}\right)\exp\left(ip_0\hat{q}\right),
\end{split} \label{eq:displacequad}
\end{equation}
one can easily reach the space representation:
%
\begin{equation}
\psi_{\alpha}(q) = \pi^{-1/4} \exp \left[ - \frac{(q-q_0)^2}{2} +ip_0q - \frac{ip_0 q_0}{2} \right]. \label{eq:psicoherent}
\end{equation}

Another formal proprieties of the coherent states turns out to be quite useful. They form a complete set,
\begin{equation}
\int_{-\infty}^{\infty}\int_{-\infty}^{\infty} \ket{\alpha}\!\bra{\alpha} \mathrm{d}q_0 \mathrm{d}p_0. = \mathbb{1},
\end{equation}	
that is, in the sense that we may express physical quantities in a coherent-state basis. Indeed, they form an over-complete set because fewer of them form a basis already\footnote{In fact, the propriety of been over-complete is a side of their lack of strict orthogonality.}, hence they are not orthogonal\footnote{As been said before, they are not eigenstates of a Hermitian operator.} and do overlap. 

For the last but not less important, the coherent states are states of minimal uncertainty in the sense that they saturate Heisenberg’s inequality \cite{bengtsson2017geometry, leonhardt2005measuring}:
\begin{equation}
\Delta q \Delta p = \frac{\hbar}{2}, \label{eq:incertcoherent}
\end{equation}
with $\Delta q $ equal to $\Delta p$.

\subsection{Squeezed States}
\begin{figure}
  \includegraphics[width=\linewidth]{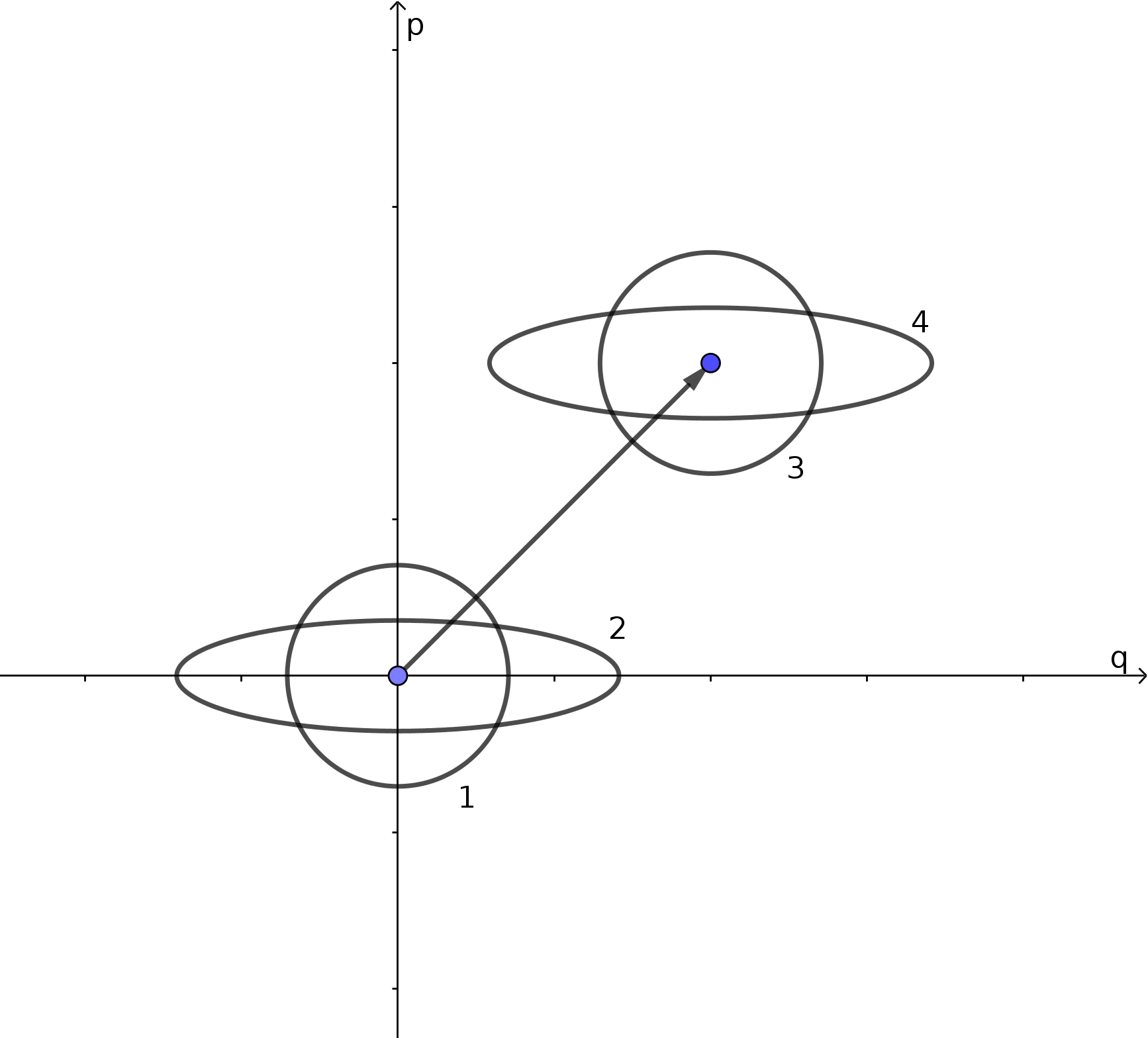}
  \caption{Comparison of different variance shape for displaced, squeezed and vacuum states. The circles 1 and 3 are vacuum and displaced vacuum state errors, respectively. And the ellipses 2 and 4 are squeezed vacuum and displaced squeezed vacuum state errors, respectively.
}
  \label{fig:squeezedstates}
\end{figure}

There is another class of states that also saturate the uncertainty relations described on eq. (\ref{eq:incertcoherent}), alongside the coherent states.
The \textit{squeezed states} maintain this property allowing unbalanced variances on the two canonical quadratures for each mode. For instance, the statistical uncertainty on position may be \textit{squeezed}, \ie, small variance, at the cost of enhancing the corresponding uncertainty on momentum and vice versa. The variance shape of such states is shown on Gif \ref{fig:squeezedstates}.

Single-mode squeezing occurs under the action of operator:
\begin{equation}
\hat{S}(\zeta)= \exp \left[\frac{\zeta}{2}\left(\hat{a}^2-\hat{a}^{\dagger 2}\right)\right],
\end{equation}
with $\zeta$ may be a complex number called \textit{squeezing parameter}. The simplest single mode squeezed state is the squeezed vacuum state,

\begin{equation}
\ket{\zeta,0}=\hat{S}(\zeta)\ket{0}.
\end{equation}

Squeezed light can be generated from light in a coherent state or vacuum state by using certain optical nonlinear interactions. According to Pauli's proof (\cite{pauli2012general}), he conjectured that states with minimum uncertainty are displaced squeezed vacuums,
\begin{equation}
\ket{\psi}=\hat{D}(\alpha)\hat{S}(\zeta)\ket{0},
\end{equation}
having the position function:
\begin{equation}
\psi(q)=\frac{e^{\zeta/2}}{\pi^{1/4}} \exp \left[-e^{2\zeta}\frac{(q-q_0)^2}{2}+ipq-\frac{ip_0q_0}{2}\right].
\end{equation}
This is the most general Gaussian pure state of a single mode.

\section{Homodyne Detection}

Now that we know how to write some continuous variable states, how about sampling it on the laboratory? More specifically, how to measure the quadratures?

First, let us introduce the \textit{phase-shift} operator

\begin{equation}
\hat{U}(\theta) = \exp(-i\theta\hat{n}). 
\end{equation}
As the name suggest, it provides the amplitude $\hat{a}$ with a phase shift $\theta$ when acting on $\hat{a}$
\begin{equation}
\hat{U}^{\dagger}(\theta)\hat{a}\hat{U}(\theta)=\hat{a}\exp(-i\theta).
\end{equation}
Form this, one can write the rotated quadrature operators to a certain reference phase $\theta$:
\begin{align}
\hat{q}_{\theta} &= \hat{q} \cos{\theta} + \hat{p}\sin{\theta} \label{eq:rotq} \\ 
\hat{p}_{\theta} &= - \hat{q}\sin{\theta} + \hat{p} \cos{\theta} 
\end{align}

\begin{figure}
  \includegraphics[width=\linewidth]{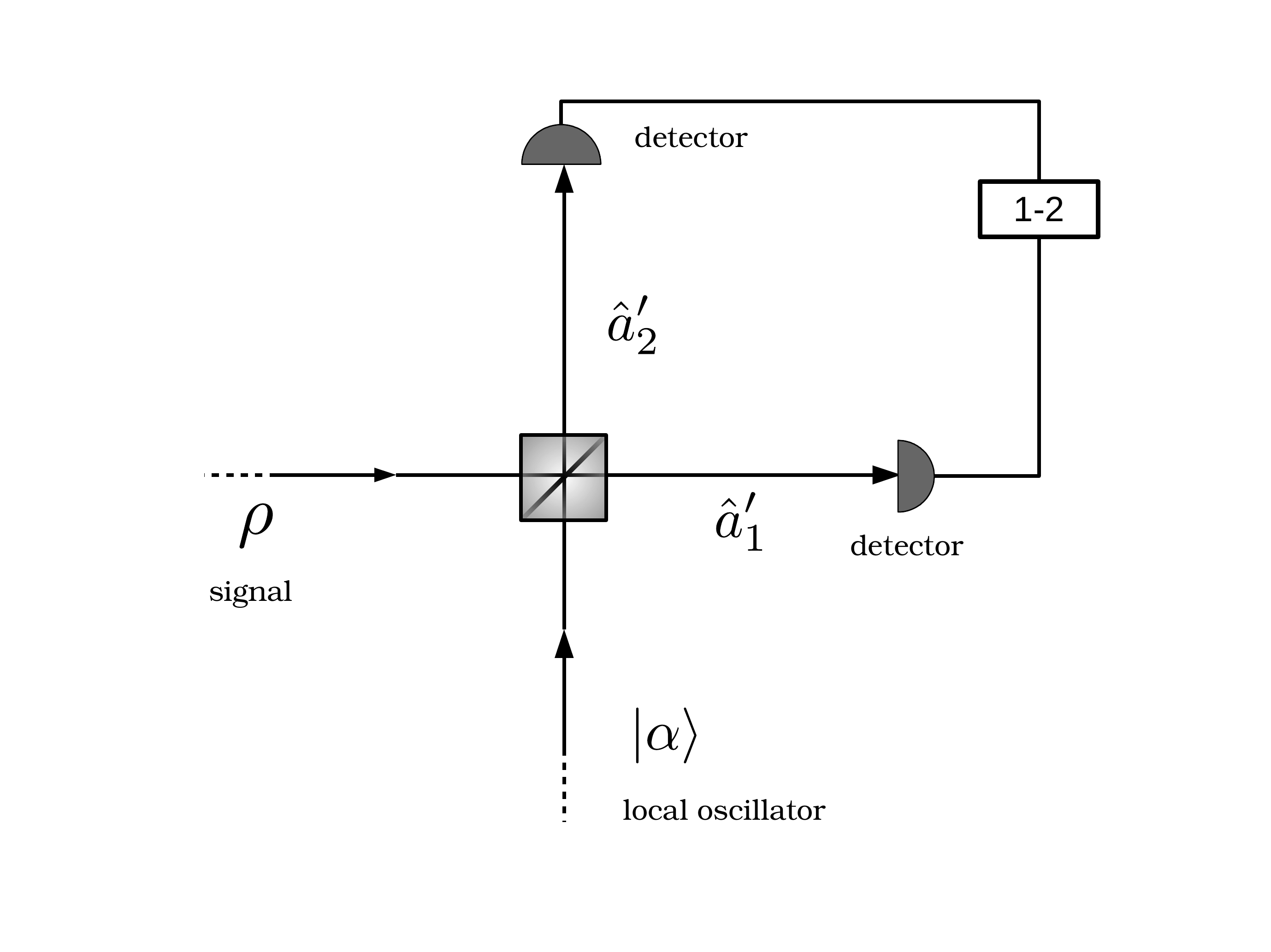}
  \vspace{-30pt}
  \caption{Diagram of a balanced homodyne tomography experiment}  
  \label{fig:homodetector}
\end{figure}

Considering that the reference phase can be varied experimentally \footnote{It means that we can go, for example, from a position representation to a momentum representation via phase shift $\theta$ of $\pi/2$}, let us make use of the usual scheme of the balanced \textit{homodyne detector}. The signal interferes with a coherent laser beam at a well-balanced 50:50 beam splitter. The scheme is on the figure \ref{fig:homodetector}. The laser field is called \textit{local oscilator} (LO). It provides the phase reference $\theta$ for the quadrature measurement. After optical mixing of the signal with the local oscillator, each emerging beam is directed to a photon detector. The photocurrents $I_1$ and $I_2$ are measured and subtracted from each other. The differences $I_{21} = I_2 - I_1$ is the quantity of interest because it contains the interference term of LO and the signal. We assume for simplicity that the measured photocurrents $I_1$ and $I_2$ are proportional to the photon numbers $\hat{n}_1$ and $\hat{n}_2$ of the beam striking each detector. They are given by:
\begin{equation}
\hat{n}_1 = \hat{a}_1'^{\dagger} \hat{a}_1', \quad \text{and} \quad \hat{n}_2 = \hat{a}_2'^{\dagger} \hat{a}_2' .
\end{equation}

Using the beam splitter Hamiltonian\cite{leonhardt2005measuring}, we can write the mode operators of the field emerging from the beam splitter,:
\begin{equation}
\hat{a}_1' = 2^{-1/2}(\hat{a} - \hat{a}_{LO}), \quad \hat{a}_2' = 2^{-1/2}(\hat{a} + \hat{a}_{LO}), 
\end{equation}
the $\hat{a}$ and $\hat{a}_{LO}$ are the annihilator operator for the signal and the local oscillator, respectively. The difference $I_{21}$ is proportional to the difference photon number\footnote{Assuming perfect quantum efficiency.}
\begin{equation}
\hat{n}_{21} = \hat{n}_2 - \hat{n}_1 = \hat{a}_{LO}^{\dagger} \hat{a} + \hat{a}_{LO}\hat{a}^{\dagger}. \label{eq:n21}
\end{equation}
We will assume that the LO is powerful enough to be treated classically, then we substitute $\hat{a}_{LO}$ by the complex amplitude $\alpha_{LO}$ and denote the phase of the local oscillator by $\theta$. Writing it in the polar form:
\begin{equation}
\alpha = \abs{\alpha_{LO}}(\cos{\theta} +i\sin{\theta}),
\end{equation}
and using the reverse relation for the creation and annihilation operators on (\ref{eq:anihhcreation}), equation (\ref{eq:n21}) is now
\begin{equation}
\hat{n}_{21} = \frac{1}{2} \abs{\alpha_{LO}} \left[ (\cos{\theta} -i\sin{\theta})(\hat{q} + i\hat{p}) + (\cos{\theta} +i\sin{\theta})(\hat{q} - i\hat{p}) \right].
\end{equation}
From the definition of the rotated quadratures (\ref{eq:rotq}), we have then
\begin{equation}
\hat{n}_{21} = \frac{1}{\sqrt{2}}\abs{\alpha_{LO}}\hat{q}_{\theta}.
\end{equation}
Therefore, a balanced homodyne detector measures the quadrature operator $\hat{q}_{\theta}$.

Now, we have the following theorem \cite{Bertrand1987}:
\begin{teo}[Bertrand and Bertrand's]
The function $W(q,p))$ is uniquely determined by the requirement that:
\begin{equation}
\bra{q_{\theta}}\rho \ket{q_{\theta}} = \frac{1}{2\pi\hbar} \int_{-\infty}^{\infty}  W(q_{\theta}\cos{\theta} - p_{\theta}\sin{\theta},q_\theta \sin{\theta} + p_{\theta}\cos{\theta}) \mathrm{d}p_{\theta} \label{eq:radontrans}
\end{equation}
for all values of $\theta$.
\end{teo}

\begin{figure}
  \includegraphics[width=\linewidth]{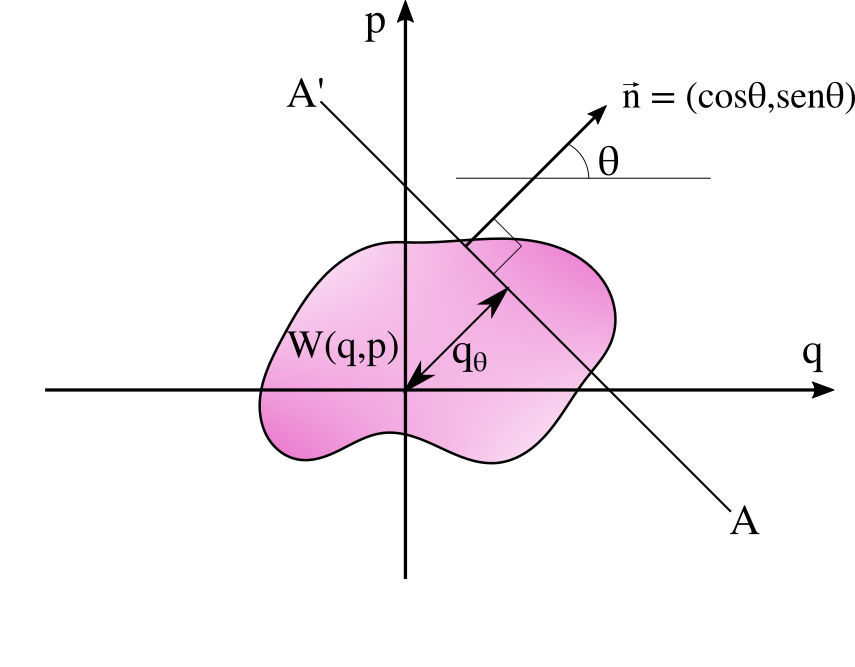}
  \vspace{-30pt}
  \caption{The Radon transform $\mathrm{pr}(q,\theta)$ of the function $W(q,p)$
is found by integrating the function along the line connecting
$A$ and $A'$}
  \label{fig:radonacting}
\end{figure}

\begin{figure}
  \includegraphics[width=\linewidth]{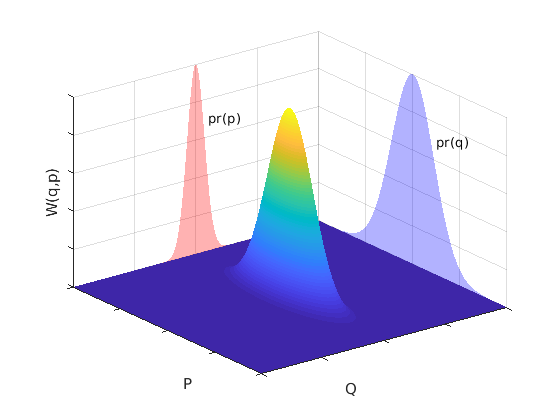}
  \caption{In homodyne tomography the Wigner function $W(q,p)$ plays the role of the unknown object. The observable ``quantum shadows'' are the quadrature distribution. In this figure, we can see the quadratures marginals $\mathrm{pr}(q)=\bra{q}\rho \ket{q}$ and $\mathrm{pr}(p)=\bra{p}\rho \ket{p}$. From the general quadrature operator $q_\theta$ distributions, the Wigner function or, more generally, the quantum state is reconstructed.}
  \label{fig:radomobral}
\end{figure}
It means the statistical distribution of the measured rotated quadrature $\hat{q}_{\theta}$ operator will equal the \textit{Radon Transform}\footnote{The relation between the operator $q_\theta$ and the Radon transform will be better explained in next chapter.}, of the Wigner Function as we can see on the Fig. \ref{fig:radomobral}. Let us see for example, the action of the Radon transform over the Gaussian Wigner function $W_G(q,p) = 1/\pi \exp (-q^2 - p^2)$:

\begin{gather*}
\frac{1}{\pi}\int_{-\infty}^{\infty} \exp[-{(q_\theta\cos{\theta}-p_\theta\sin{\theta})}^2]\exp[-{(q_\theta\sin{\theta}+p_\theta\cos{\theta})}^2] \mathrm{d}p_\theta \\
\frac{1}{\pi}\int_{-\infty}^{\infty} \exp[-q_\theta^2\cos^2{\theta} -q_\theta^2\sin^2{\theta}] \exp[-p_\theta^2\sin^2{\theta} -p_\theta^2\cos^2{\theta}] \mathrm{d}p_\theta \\
\frac{1}{\sqrt{\pi}}\exp[-q_\theta^2]= \frac{1}{\pi^{1/4}}\exp\left[\frac{-q_\theta^2}{2}\right] \times \frac{1}{\pi^{1/4}}\exp\left[\frac{-q_\theta^2}{2}\right] = \braket{q_\theta}{\psi_G}\braket{\psi_G}{q_\theta}.
\end{gather*}

As one can see, the Radon transform indeed returns the marginal distribution over the quadrature operator $q_\theta$.

\section{Inverse Radon Transform}

Once we have measured the rotated quadrature operator $q_{\theta}$ through homodyne detection, it is intuitive to think that inverting the Radon transform (\ref{eq:radontrans}) is a good way to obtain the Wigner function, and then, the density operator $\rho$. It looks the most ``natural'' solution if we were inverting a linear system. However inversion problems are not always an easy task.

Let us check the inversion for the reconstruction of the Wigner Function. We perform a position Fourier transform on the probability distribution $pr(q,\theta) = \bra{q_{\theta}}\rho\ket{q_{\theta}} =\bra{q} \hat{U}_\theta \rho \hat{U}^{\dagger}_\theta \ket{q}$:
\begin{equation}
\begin{split}
\widetilde{pr}(\xi,\theta) &=  \int_{-\infty}^{\infty} pr(q,\theta) e^{-i\xi q/\hbar} \mathrm{d}q \\
& = \int_{-\infty}^{\infty} \bra{q} \hat{U}_\theta \rho \hat{U}_\theta^{\dagger} \ket{q} e^{-i\xi q/\hbar} \mathrm{d}q \\
& = \int_{-\infty}^{\infty} \bra{q}\hat{U}_\theta \rho \hat{U}_\theta^{\dagger} e^{-i\xi \hat{q}/\hbar} \ket{q} \mathrm{d}q \\
& = \tr [\hat{U}_\theta \rho \hat{U}_\theta^{\dagger} e^{-i\xi \hat{q}/\hbar}] = Tr [\rho \hat{U}_\theta^{\dagger} e^{-i(\xi \hat{q})/\hbar} \hat{U}_\theta].
\end{split}
\end{equation}
From the definitions of rotated quadrature operator $q_{\theta}$ (\ref{eq:rotq}) and characteristic function (\ref{eq:charfun}), we have
\begin{equation}
\widetilde{pr}(\xi,\theta) = Tr \left\{\rho e^{-i[\hat{q} \xi \cos{}\theta + \hat{p} \xi \sin{\theta}]/\hbar} \right\} = \widetilde{W}(\xi\cos{\theta},\xi\sin{\theta}). \label{eq:polarchar}
\end{equation}
In other words, the Fourier-transformed position probability distribution is the characteristic function in polar coordinates. From (\ref{eq:probabilitydensfun}), the Wigner function is a Fourier transform of the characteristic function. Performing the appropriate transforms for polar coordinates, we obtain:
\begin{equation}
\begin{split}
W(q,p) = & \frac{1}{(2\pi\hbar)^2}\int_{-\infty}^{+\infty} \int_{0}^{\pi} \widetilde{W}(\xi\cos{\theta},\xi\sin{\theta})  \\
& \times \exp{[i\xi(q\cos{\theta} + p\sin{\theta})/\hbar]} \mathrm{d}\theta \mathrm{d}\xi \\
 = &\frac{1}{(2\pi\hbar)^2} \int_{-\infty}^{+\infty} \int_{0}^{\pi} \int_{-\infty}^{+\infty} pr(x,\theta) \abs{\xi} \\
& \times \exp{[i\xi(q\cos{\theta} + p\sin{\theta} -x)/\hbar]} \mathrm{d}x \mathrm{d}\theta  \mathrm{d}\xi, \label{eq:radontranstrool}
\end{split}
\end{equation}
using (\ref{eq:polarchar}).To simplify (\ref{eq:radontranstrool}), we introduce the kernel
\begin{equation}
K(x)= \frac{1}{2} \int_{-\infty}^{+\infty} \abs{\xi} \exp{(i\xi x)} \mathrm{d}\xi , \label{eq:kernel}
\end{equation}   
and obtain
\begin{equation}
W(q,p) = \frac{1}{2 \pi^{2}} \int_{-\infty}^{+\infty} \int_{0}^{\pi} pr(x,\theta) K(q\cos{\theta} + p\sin{}\theta -x) \mathrm{d}x \mathrm{d}\theta . \label{eq:radontrue}
\end{equation}

To use the equation above, in practice we need to regularize $K(x)$. The direct demonstration of the formula is mathematically delicate, and can be found for instance in Radon’s article \cite{radontransform} and on \cite{leonhardt2005measuring}. The compact formula for the \textit{inverse Radon Transform} is
\begin{equation}
W(q,p) = -\frac{\mathcal{P}}{2\pi^2} \int_{0}^{\pi} \int_{-\infty}^{+\infty} \frac{pr(x,\theta) \mathrm{d}x \mathrm{d}\theta}{(q\cos{\theta} + p\sin{\theta}- x)^2} .
\end{equation}
where $\mathcal{P}$ is the principal-value operator of the kernel (\ref{eq:kernel}).

Although exact, this expression is nevertheless unusable with experimental data as the algebraic expression of $p(x,\theta)$ would be unknown and it would therefore be impossible to evaluate precisely the principal value of the integral. 

In the real world, it is better to regularize and replace the kernel $K(x)$ with some numerical approximation. This is possible setting a frequency cutoff $k_c$ in the definition (\ref{eq:kernel}) of the kernel $K(x)$. Which lead us to the algorithm of \textit{filtered back-projection}, a common protocol for image reconstruction.

In this case, we obtain the integral
\begin{equation}
K(x) = \frac{1}{2} \int_{-k_c}^{+k_c} \abs{\xi} e^{i\xi x }\mathrm{d}\xi,
\end{equation}
which is calculated to yield
\begin{equation}
K(x) \approx \frac{1}{x^2}[\cos{(k_c x)} + k_c\sin{(k_c x) -1}.
\end{equation}

\begin{figure}
  \includegraphics[width=\linewidth]{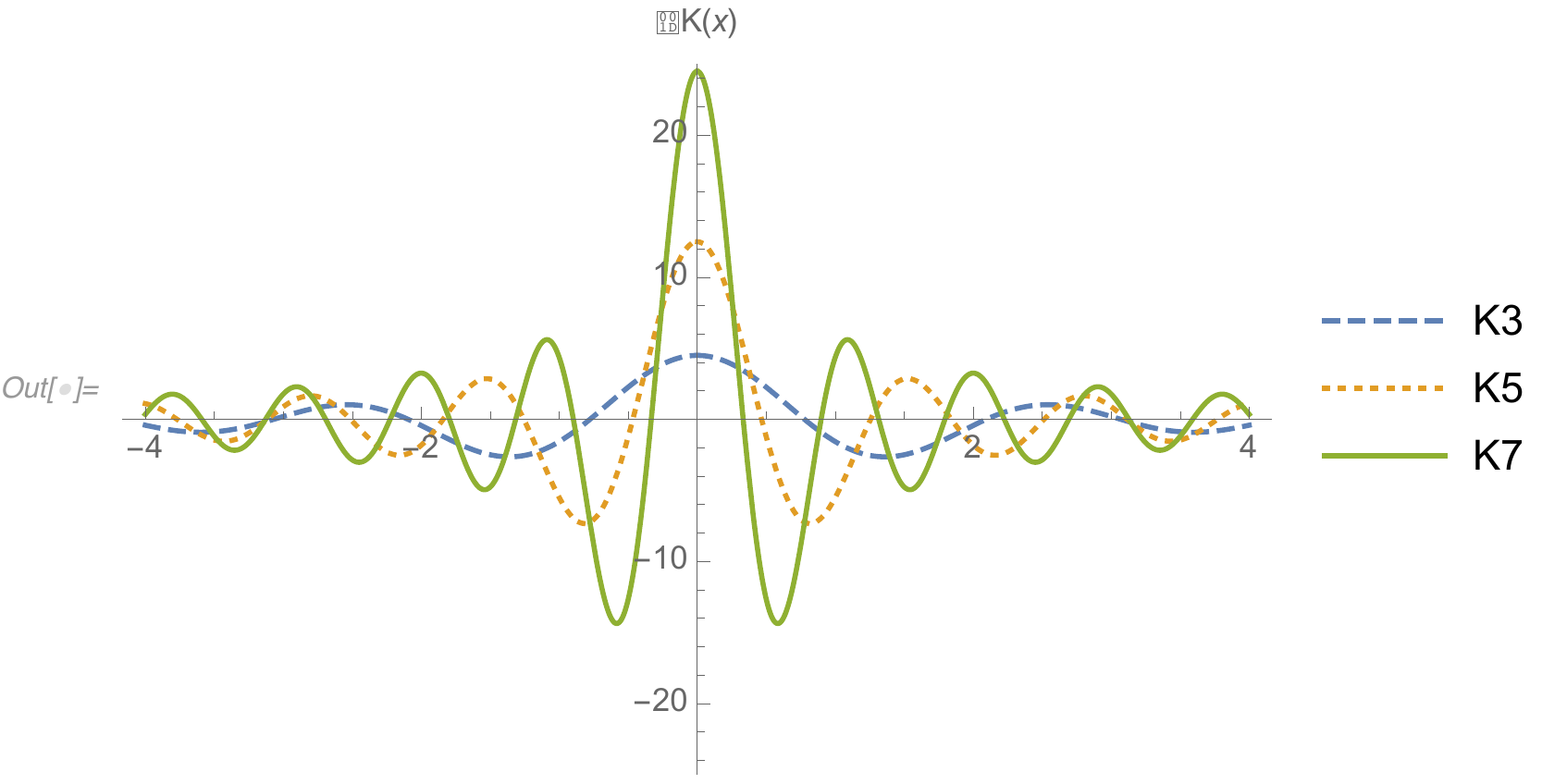}
  \caption{The approximate kernel $K(x)$ for different values of $x$.}
  \label{fig:cuttof}
\end{figure}
In practice, the choice of $k_c$ affects how detailed Wigner function will get reconstructed. For instance, choosing a low $k_c$ on the convolution in (\ref{eq:radontrue}) will filter out the fine physical details of the Wigner function. On the other hand, If $k_c$ is set too high, it will introduce nonphysical rapid oscillation noise from the statistical errors in the measurement of $p(x,\theta)$\cite{leonhardt2005measuring,benichi2011}. Choosing the right cutoff is particular for each state to be reconstructed, once you have to choose between two regimes. For instance, some cutoffs are sampled at Fig. (\ref{fig:cuttof}).

Let us think about a multi-mode inverse Radon inverse. First, we define the vectors:
\begin{equation}
\Theta = [\theta_1,...,\theta_n]^T, \quad  p_\theta = [p_{1\theta_1},...,p_{n\theta_n}]^T  \quad \hat{U}(\Theta) = \hat{U}_1(\theta_1)\otimes \dots \otimes \hat{U}_n(\theta_n) .
\end{equation}

For a multi-mode, the probability distribution has the form:
\begin{equation}
\begin{split}
pr\left(q_1, \theta_1,..., q_n, \theta_n \right) &= \exval{q_{1},..., q_{n}}{\hat{U}(\Theta)\rho \hat{U}^{\dagger}(\Theta)}{q_{1},..., q_{n}} \\
&= \int_{\infty}^{\infty} W (RO_\theta) \mathrm{d}^np_\theta
\end{split}
\end{equation}
where
\begin{equation}
O_\theta = [q_{1\theta_1},p_{1\theta_1},...,q_{n\theta_n},p_{n\theta_n}]^T,
\end{equation}
and $R$ is the sympletic matrix:
\begin{equation}\label{eq:Rzão}
R = \bigoplus_{i=1}^n \begin{bmatrix} 
\cos{\theta_i} & -\sin{\theta_i}\\
\sin{\theta_i} & \cos{\theta_i}
\end{bmatrix}.
\end{equation}
In the same fashion of the single mode derivation, we make use of the Fourier transform:
\begin{equation}
\widetilde{pr}(\xi_1, \theta_1,...,\xi_n,\theta_n) =  \int pr\left(q_1, \theta_1,..., q_n, \theta_n \right) e^{-i(\xi^Tq)/\hbar} \mathrm{d}^{n}q,
\end{equation}
with the vectors $\xi$ and $q$:
\begin{equation}
\xi = [ \xi_1,..., \xi_n ]^T , \qquad  q  = [q_1,...,q_n]^T.
\end{equation}
Then we have 
\begin{equation}
\begin{split}
\widetilde{pr}(\xi_1, \theta_1,..., \xi_n, \theta_n)  = & \int \bra{q_1,..., q_n}{\hat{U}(\Theta) \rho \hat{U}^{\dagger}(\Theta)  } e^{-i(\xi^T\hat{q})/\hbar}\ket{q_1,...,q_n} \mathrm{d}^nq \\
= & Tr\left[\hat{U}(\Theta) \rho \hat{U}^{\dagger}(\Theta) e^{-i(\xi^T\hat{q})/\hbar} \right]\\
= & Tr\left[\rho \hat{U}^{\dagger}(\Theta)  e^{-i(\xi^T\hat{q})/\hbar} \hat{U}(\Theta) \right] \\
= & Tr\left[\rho e^{-i[\xi_1(q_1\cos{\theta_1}+p_1\sin{\theta_1})+...+\xi_n(q_n\cos{\theta_n}+p_n\sin{\theta_n})]/\hbar}\right] \\
= & \widetilde{W}(\xi_1\cos{\theta_1},\xi_1\sin{\theta_1},..., \xi_n\cos{\theta_n}, \xi_n\sin{\theta_n}) \\
\end{split}.
\end{equation}

Now we can set our multi-mode Wigner function $W(O)$, with $O$ as the measured values of the operator $\hat{O}$  (\ref{eq:vectoroperator}):
\begin{equation}
\begin{split}
W(O) = & \frac{1}{(2\pi \hbar)^{2n}}\int  \widetilde{W}(\xi_1\cos{\theta_1},\xi_1\sin{\theta_1},..., \xi_n\cos{\theta_n}, \xi_n\sin{\theta_n})  \mathrm{d}^n \xi \mathrm{d}^n \Theta \\
=& \frac{1}{(2\pi \hbar)^{2n}}\int  pr(x_1,\theta_1,...,x_n,\theta_n) \abs{\xi_1}\exp\left\{i[(\xi_1(q_1\cos{\theta_1}+p_1\sin{\theta_1} -x_1)]/\hbar\right\} \times \dots \\
& \dots \times \abs{\xi_n}\exp\left\{i[(\xi_n(q_n\cos{\theta_n}+p_n\sin{\theta_n} - x_n)]/\hbar\right\}\mathrm{d}^n x \mathrm{d}^n \xi \mathrm{d}^n \Theta.
\end{split}
\end{equation}
As we can see, for each mode now we have a different kernel, defined in (\ref{eq:kernel}), so we can compute the equation above as
\begin{equation}
\begin{split}
W(O) = & \frac{1}{(2\pi \hbar)^{2n}}\int  pr(x_1,\theta_1,...,x_n,\theta_n) K(q_1\cos{\theta_1}+p_1\sin{\theta_1}-x_1)\times \\ 
& \dots \times K(q_n\cos{\theta_n}+p_n\sin{\theta_n}-x_n) \mathrm{d}^n x \mathrm{d}^n \Theta.
\end{split}
\end{equation}

\begin{figure}
\centering
\includegraphics[width=\textwidth]{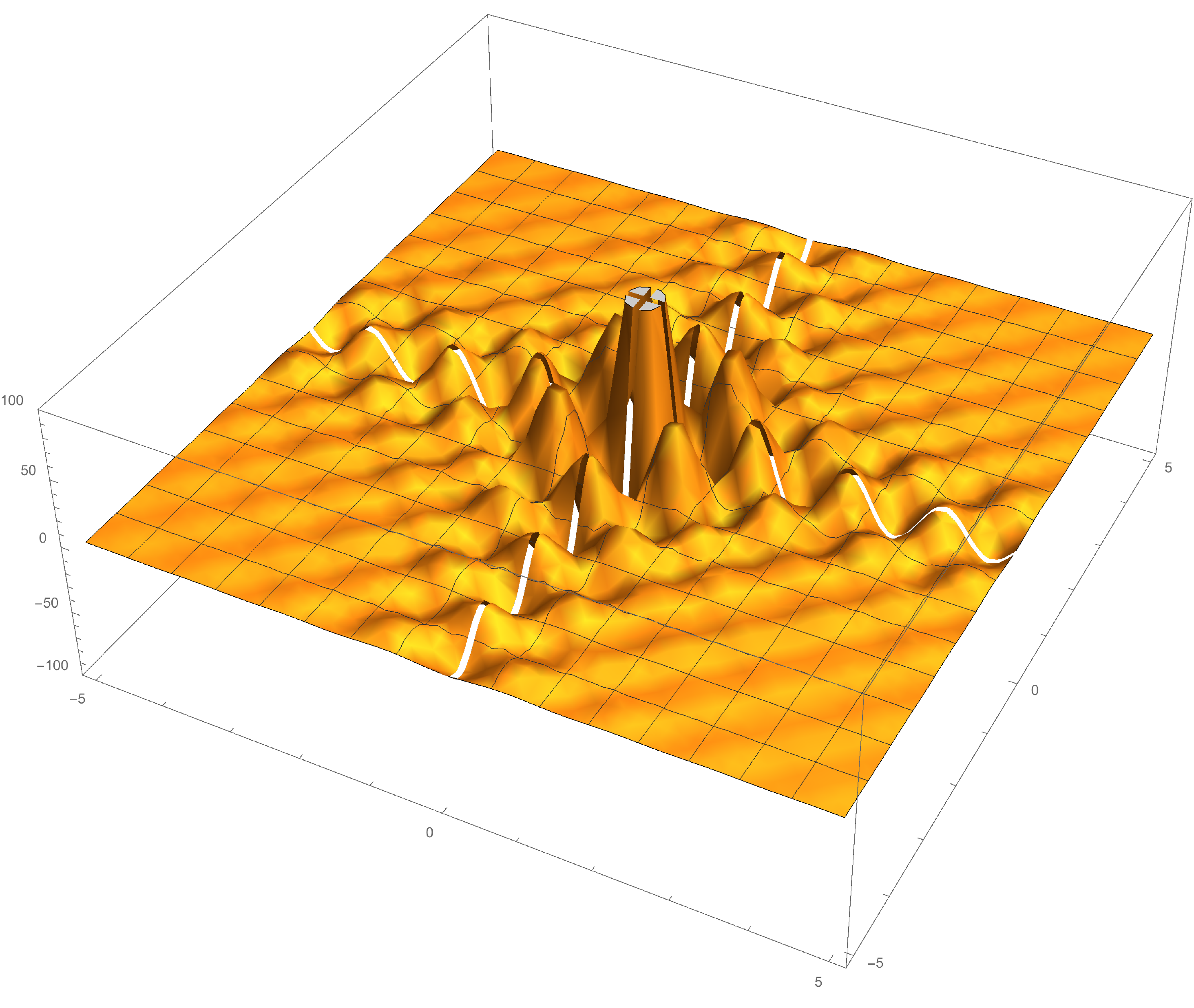}
  \caption{Kernels of a two mode state heavily interfere with each other.}
  \label{fig:kkkcuttof}
\end{figure} 

To visualize and understand it better, the product of two mode kernels are shown figure (\ref{fig:kkkcuttof}). As we can see, the kernel values for each individual mode interfere with all the others, therefore causing the noise to increase very fast for a many mode state. Which means it is a hard task to achieve a good visualization for the state function in both local or global stances. The coast to adjust cutoffs and the intrinsically error from the measurement makes back-projection algorithm for many modes inefficient.

\chapter{The Toolbox: Design for Zeros and Ones}

\epigraphfontsize{\small\itshape}
\epigraph{"Machines take me by surprise with great frequency."}{--- \textup{Alan Turing}}


In the last chapter, we learned about how to deal with continuous variable states, from the representation to the measurement process. We are now ready to perform some calculations. How about a computer to make easier our tasks? 

When we are dealing with bits there is no other way: we need to discretize our function. This happens through advanced techniques of mapping input values from a large set (often a continuous set) to output values in a (countable) smaller set, which means that information is compressed. Despite the fact of some amount of it is lost in the process, we can still get precise information about the state.


To build the algorithms and perform the calculations, on this dissertation we used MATLAB environment, since it has the necessary libraries to compute matrices, to use linear algebra, numerical integration and do simple symbolic functions. We also make use of Wolfram Mathematica to work with more complicate analytical functions.

In this chapter, we investigate the visualization of the states through Wigner Functions, the simulation of a homodyne measurement and the possibility of projecting the state in the Fock basis, which leads us to truncate the density state operator in a sufficient accurate matrix representation. Besides, we discuss briefly on how to improve the reconstructed states through semidefinite programming.

\section{Samples of Wigner Functions}

The transformation of the state probability density function from a space or momentum representation is linear through the Wigner function (\ref{eq:wignerfunction}). Our script generates a matrix with adjustable resolution.\footnote{On this dissertation, the standard resolution is $100\times100$ points, and the ranges are $-5$ to $5$ to the $q$ and $p$ axes. Also, we have set $\hbar = 1$ to make easier the writing of the scripts.} Given an input in position representation\footnote{We gave preference for position representation to write the input state, since the momentum representation is the Fourier transform of the previous one.} $\psi(x)$, the output is a two dimensional array $W(q,p)$, which can be used to generate a plot, as we can see on the next pages. The reverse process, considering a pure state, from the Wigner Function to the position state representation, we can invert the relation (\ref{eq:wignerfunction}):
\begin{equation}
\bra{x'}\rho\ket{x'} = \psi(x)\psi^*(x') = \int W\left(\frac{x+x'}{2},p\right)e^{-ip(x-x')/\hbar}\mathrm{d}p,
\end{equation}
and setting $\psi^*(x')$ for $x=0$:
\begin{equation}
\psi(x) = \frac{1}{\psi^*(0)} \int W\left(\frac{x}{2},p\right)e^{-ipx/\hbar} \mathrm{d}p.
\end{equation}
Here $\psi^*(0)$ acts as normalization factor and we obtain the original position representation to the state. 

\begin{figure}
    \centering
    \begin{subfigure}{\textwidth}
        \includegraphics[width=\textwidth]{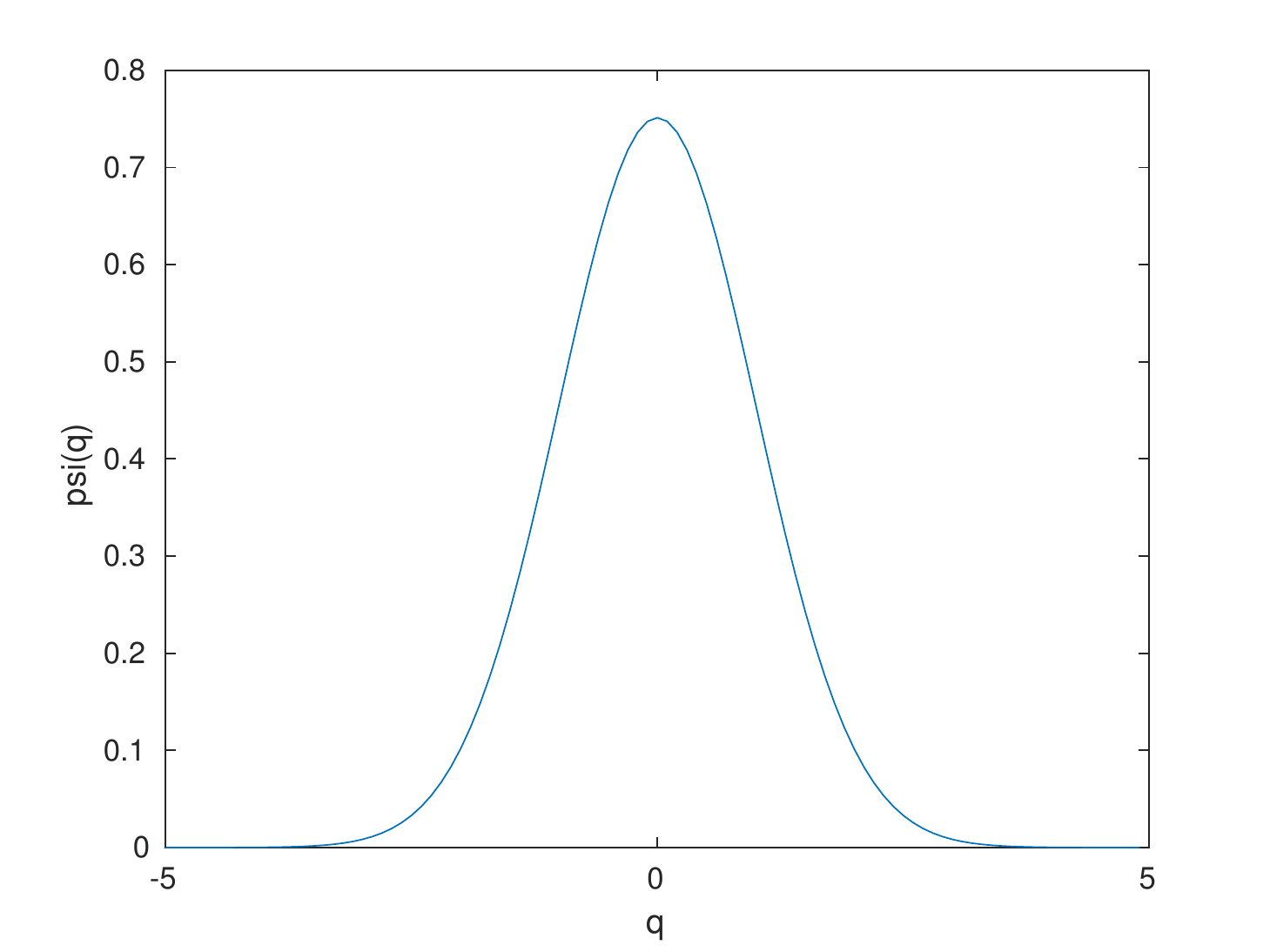}
        \centering
        \caption{Position representation of the vacuum state $\ket{0}$}
        \label{fig:psivac}
    \end{subfigure}
    ~ 
    \begin{subfigure}{\textwidth}
    	\centering
        \includegraphics[width=\textwidth]{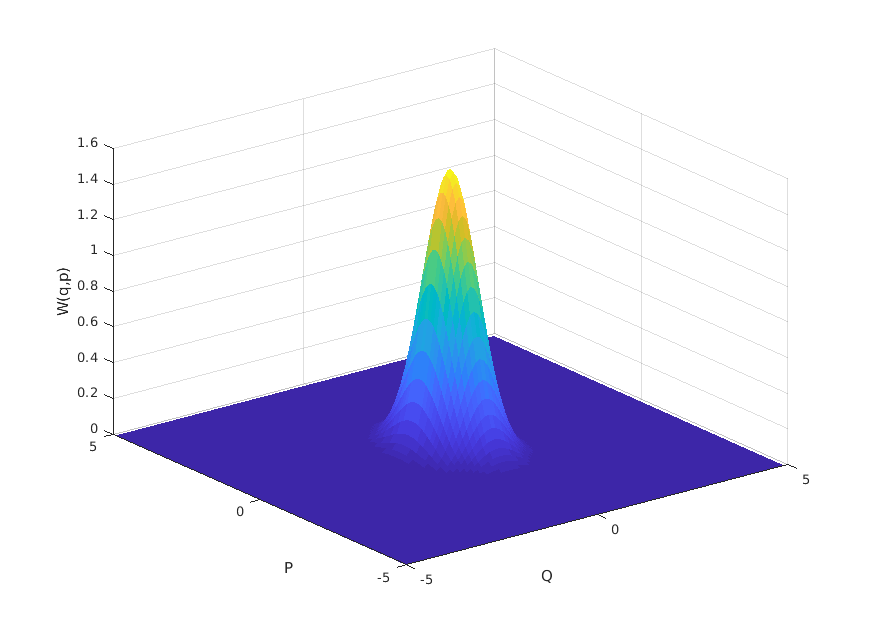}
        \caption{Wigner Function of vacuum state}
        \label{fig:wigvac}
    \end{subfigure}
    \caption{Diferent ways to visualize the vacuum state.}\label{fig:vacuumrepre}
\end{figure}

The simplest case is the vacuum state (\ref{eq:psivacuum}): a Gaussian function leading to another Gaussian function. We have:
\begin{equation}
W_{vacuum}(q,p) = \frac{1}{\pi\hbar}\exp\left(\frac{-q^2}{2}\right) \exp\left(\frac{-2p^2}{\hbar^2}\right).
\end{equation}
We can see the representation of the functions for vacuum state on figure (\ref{fig:vacuumrepre}).


\subsection{Coherent States}
Since the coherent sates are displaced vacuum, we are induced to think that corresponding Wigner functions are displaced vacuum Wigner functions too, with the displacement given by the complex amplitude $\sqrt{2} \alpha = q_0 + ip_0$. Using the displacement operator $\hat{D}$ in quadrature representation (\ref{eq:displacequad}):
\begin{equation}
\begin{split}
W_D(q,p) = &\frac{1}{2\pi\hbar}\int_{-\infty}^{+\infty} \bra{q-\frac{x}{2}}\hat{D}\rho\hat{D}^{\dagger}\ket{q+\frac{x}{2}} e^{ipx/\hbar}\mathrm{d}x\\
= & \frac{1}{2\pi\hbar}\int_{-\infty}^{+\infty} \bra{q-\frac{x}{2}}e^{-iq_0\hat{p}}e^{ip_0\hat{q}}\rho e^{-ip_0\hat{q}}e^{iq_0\hat{p}}\ket{q+\frac{x}{2}} e^{ipx/\hbar}\mathrm{d}x\\
= &  \frac{1}{2\pi\hbar}\int_{-\infty}^{+\infty} \bra{q-\frac{x}{2}-q_0}e^{ip_0\hat{q}}\rho e^{-ip_0\hat{q}}\ket{q+\frac{x}{2}-q_0} e^{ipx/\hbar}\mathrm{d}x\\
=& \frac{1}{2\pi\hbar}\int_{-\infty}^{+\infty} \bra{q-\frac{x}{2}-q_0}\rho\ket{q+\frac{x}{2}-q_0}e^{i(p-p_0)x/\hbar}.
\end{split}
\end{equation}
So it is indeed displaced Wigner functions
\begin{equation}
W_D(q,p) = W(q-q_0,p-p_0),
\end{equation}
and the Wigner function of a coherent state is given by the displaced Gaussian: 
\begin{equation}
W_{\alpha}(q,p) = \frac{1}{\pi\hbar}\exp\left[\frac{-(q-q_0)^2}{2}\right] \exp\left[\frac{-(p-p_0)^22}{\hbar^2}\right].\label{eq:wignercoherent}
\end{equation}
Since we can set the values for $\alpha$, the algorithms makes a translation on the phase space setting $q_0=\mathrm{Re}(\alpha)$ and $p_0=\mathrm{Im}(\alpha)$. Another alternative is to generate the Wigner functions doing the numerical integration of the position functions (\ref{eq:psicoherent}).

\begin{figure}
    \centering
    \begin{subfigure}{\textwidth}
        \centering
        \includegraphics[width=\textwidth]{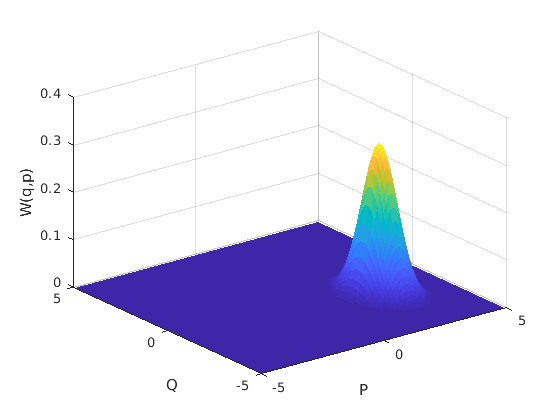}
        \caption{Displaced vacuum with $\alpha = -1.5 + i2$}
        \label{fig:displacedvacuum}
    \end{subfigure}
    ~ 
    \begin{subfigure}{\textwidth}
    	\centering
        \includegraphics[width=\textwidth]{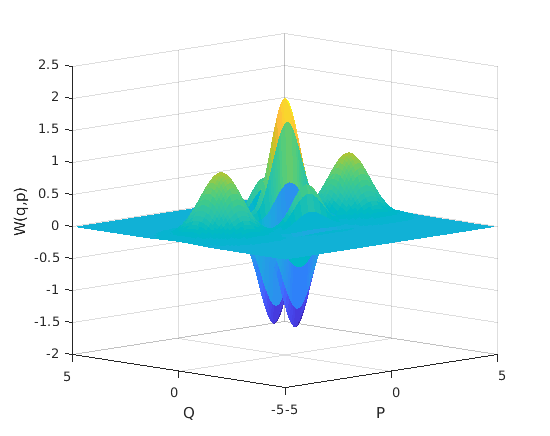}
        \caption{Schrödinger Cat for $\ket{\alpha = 3}+\ket{\alpha = -3}$}
        \label{fig:cat3}
    \end{subfigure}
    \caption{Pictures of coherent state and a superposition of coherent states, knwon as Schrödinger Cat}\label{fig:displacedandcat}
\end{figure}

If we think about the fundamental superposition principle of quantum mechanics, how a superposition of coherent states would look like? In figure \ref{fig:cat3} we show the Wigner function for a Schrödinger cat state, which is the superposition of the states $\ket{\alpha}$ and $\ket{-\alpha}$. These two states are usually taken to represent the cat's macroscopically distinguishable states $\ket{alive}$ and $\ket{dead}$ from Schrödinger's famous \textit{gedankenexperiment}\footnote{From German: "thought experiment", its used to describe  some hypothesis, theory or principle for the purpose of thinking through its consequences. Given the structure of the experiment, it may not be possible to perform it, and even if it could be performed, there need not be an intention to perform it.\cite{wikpediagedan}}. The position functions shows two peaks, one at $q_0$ and the other at $-q_0$ according to the superimposed coherent amplitudes, and between there are rapid oscillations with large negative values, indicating the nonclassical behavior of the Schrödinger cat state. The generation and quantum tomography of cat states has only been realized recently because they are extremely vulnerable to losses \cite{leonhardt2005measuring,Lvovsky2009,Braunstein2004}.

These states are useful for many quantum information protocols such as quantum teleportation \cite{vanEnkHirota2003}, quantum computation \cite{Ralph2003}, and error correction \cite{cochrane1999}. It is thus not surprising that experimental synthesis of Schrödinger cats has been an object of aspiration for several generations of physicists.

\subsection{Squeezed states}

What is the Wigner function for a squeezed state? From the Wigner formula (\ref{eq:wignerfunction}):
\begin{equation}
\begin{split}
W_s(q,p) = & \frac{1}{2\pi \hbar} \int_{-\infty}^{\infty} \bra{q-\frac{x}{2}}\hat{S} \rho hat{S}^{\dagger}\ket{q+\frac{x}{2}}e^{ipx/\hbar}\mathrm{d}x\\
= & \frac{1}{2\pi \hbar} \int_{-\infty}^{\infty} \bra{e^\zeta \left(q-\frac{x}{2}\right)} \rho \ket{e^\zeta \left(q+\frac{x}{2}\right)}e^{ipx/\hbar} e^\zeta \mathrm{d}x
\end{split}
\end{equation}
substituting the $e^{\zeta}x$ with $x'$, we get the result
\begin{equation}
W_s(q,p) = W(e^{\zeta} q, e^{-\zeta}p).
\end{equation}

In order to preserve the area in phase space, the Wigner function for a squeezed state is squeezed in one quadrature direction and stretched accordingly in the orthogonal one.

\begin{figure}
    \centering
    \begin{subfigure}{\textwidth}
        \centering
        \includegraphics[width=\textwidth]{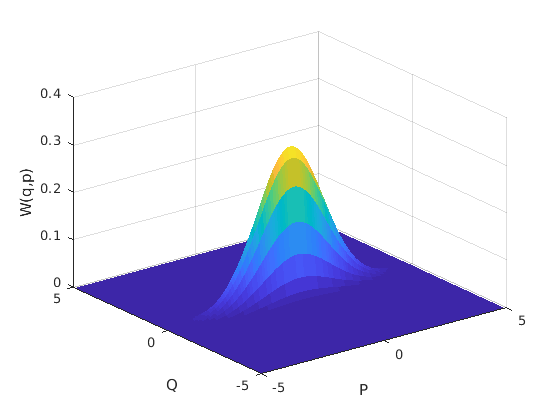}
        \caption{Squeezed vacuum}
        \label{fig:squeezedvacuum}
    \end{subfigure}
    ~ 
    \begin{subfigure}{\textwidth}
    	\centering
        \includegraphics[width=\textwidth]{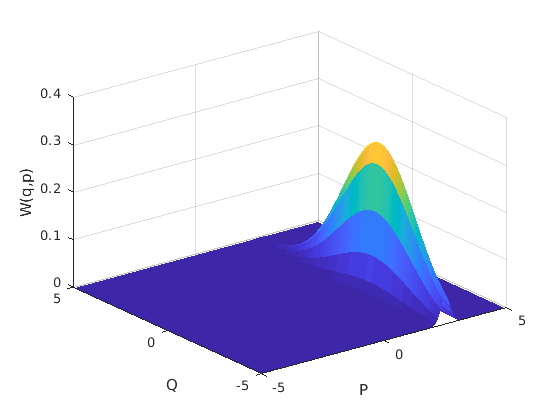}
        \caption{Displaced squeezed vacuum}
        \label{fig:displacedsqueezed}
    \end{subfigure}
    \caption{Samples of squeezed states. The figure (b) is the most general Gaussian state on Wigner representation.}
     \label{fig:coherentexemples}
\end{figure}

For instance, the Wigner function of a squeezed vacuum:
\begin{equation}
W_s(q,p) = \frac{1}{\pi\hbar} \exp \left(\frac{-e^{2\zeta}q^2}{2}\right) \exp \left(\frac{- 2e^{2\zeta}p^2}{\hbar^2}, \right)
\end{equation}
which also has Gaussian form, however, with unbalanced variances indicating the effect of quadrature squeezing. As we have been using, given the position function, the numerical Wigner functions can be found easily. Combining the displaced vacuum and the squeezed vacuum algorithms, we have made a third one that can produce the most general Wigner function for Gaussian state, given the displacement $\alpha$ and the squeezing factor $\zeta$ as inputs. The plots of the functions are available on figure \ref{fig:displacedsqueezed}. Some of the first reconstructions of Wigner functions were indeed those of squeezed states \cite{Smithey1993b,breitenbach1997measurement}.

\subsection{Fock states}

\begin{figure}
    \centering
    \begin{subfigure}{0.45\textwidth}
        \includegraphics[width=\textwidth]{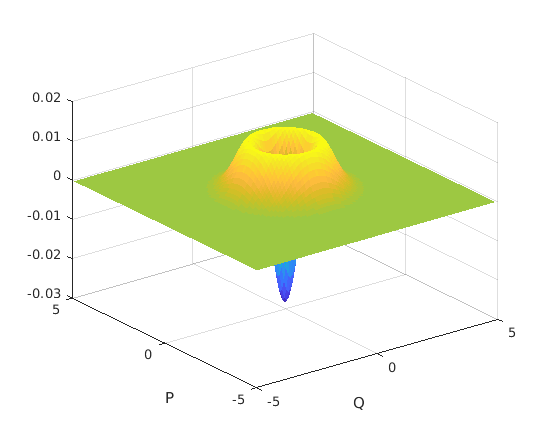}
        \caption{Fock state$\ket{1}$}
        \label{fcok1}
    \end{subfigure}
    ~ 
    \begin{subfigure}{0.45\textwidth}
        \includegraphics[width=\textwidth]{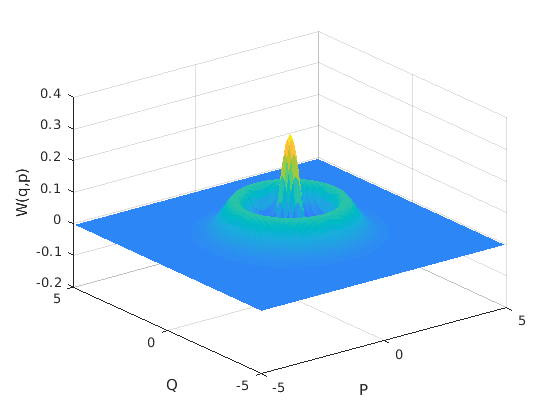}
        \caption{Fock state$\ket{2}$}
        \label{fig:fock2}
    \end{subfigure}
    ~ 
    \begin{subfigure}{0.45\textwidth}
        \includegraphics[width=\textwidth]{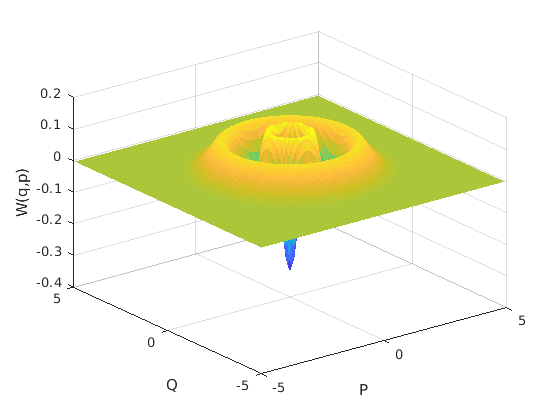}
        \caption{Fock state$\ket{3}$}
        \label{fig:fock3}
    \end{subfigure}
    ~ 
    \begin{subfigure}{0.45\textwidth}
        \includegraphics[width=\textwidth]{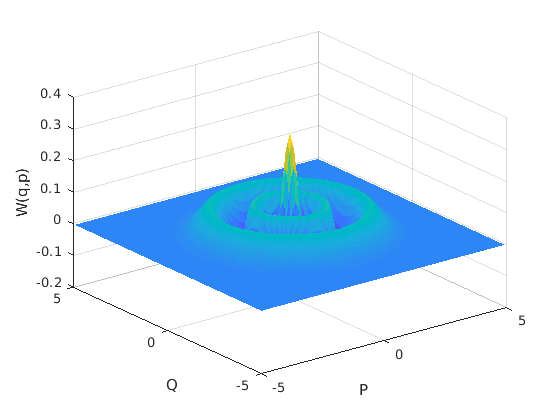}
        \caption{Fock state$\ket{4}$}
        \label{fig:fock4}
    \end{subfigure}
    \caption{Wigner functions of the Fock state from $\ket{1}$ to $\ket{4}$.}\label{fig:focksfig}
\end{figure}

The Wigner functions for the Fock states are shown in figure \ref{fig:focksfig}. Several common features are immediately apparent between these functions: they resemble the position representation for $\ket{n}$ in having $n$ zero-crossings, the functions are all radially symmetric, the even $n$ states have a peak at the origin, while the odd $n$ states have a dip at the origin. The peaks and dips have the same amplitude.

Though the eq. (\ref{fockhermite}), we can generate the $\psi_n(x)$ and use it as input on the algorithm of the numerical Wigner function. Note that in this case, we actually expanding the Hermite polynomials, which routine is already known and can be efficiently calculated.

Acording to \cite{leonhardt2005measuring}, the Wigner function $W_n(q,p)$ of Fock states is
\begin{equation}
W_n(q,p) = \frac{(-1)^n}{\pi\hbar}\exp(-q^2-p^2)L_n(2q^2+2p^2).
\end{equation}

\subsection{Two modes}

Mapping the Wigner function for more modes is trickier to visualize and calculate. For each mode, we add a new pair of parameters $(q_i, p_i)$, which means, for a two modes Wigner function, we need to store a multidimensional array $W(q_1,p_1,q_2,p_2)$. Since the calculation happens through a series of integrations, the algorithm becomes slow and inefficient. For some states, it is not even feasible using the numerical integration routine.

It is important to emphasize that in order to study \textit{entanglement}\footnote{I will discuss better about it on the next chapter.} on continuous variable states, one of the first made is a two mode squeezed vacuum \cite{leonhardt2005measuring}
\begin{equation}
\psi(q_1,q_2) = \pi^{-1/2}\exp \left[-\frac{1}{4}e^{2\zeta}(q_1+q_2)^2-\frac{1}{4}e^{-2\zeta}(q_1-q_2)^2 \right],
\end{equation}	
which describes an entangled state (with given mean energy), and it provides physical realization. 

Although the analytical treatment of multimode Gaussian states has been study already \cite{Simon1987}, the numerical reconstruction in this case still a challenge.

\section{Tomography protocol}

\begin{figure}
    \centering
    \begin{subfigure}{0.3\textwidth}
    	\centering
    	\vspace{15pt}
        \includegraphics[width=\textwidth]{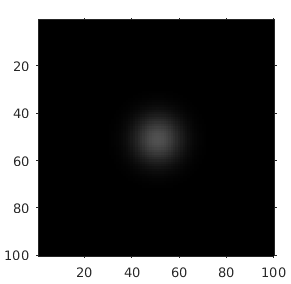}
        \vspace{7pt}
        \caption{Vacuum state as image in gray scale conversion}
        \label{fig:grayvac}
    \end{subfigure}
    ~ 
    \begin{subfigure}{0.55\textwidth}
        \centering
        \includegraphics[width=\textwidth]{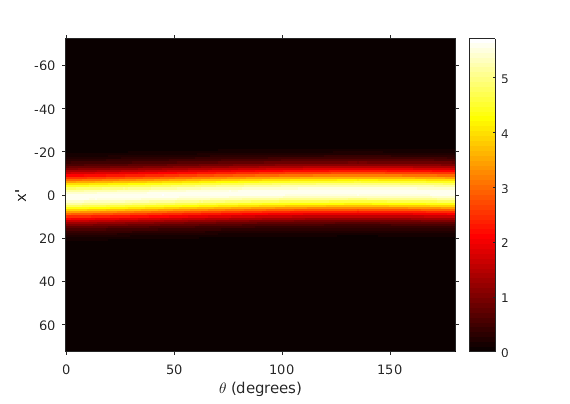}
        \caption{The Radon transform, with $\theta$ ranging from 0 to 180 in integer steps of size 1.}
        \label{fig:radonvac}
    \end{subfigure}
    \caption{Representation of the vacuum state as image in greyscale and the correspounding Radon transform.}\label{fig:radontrans}
\end{figure}

\begin{figure}   
    \begin{subfigure}{0.45\textwidth}
        \centering
        \includegraphics[width=\textwidth]{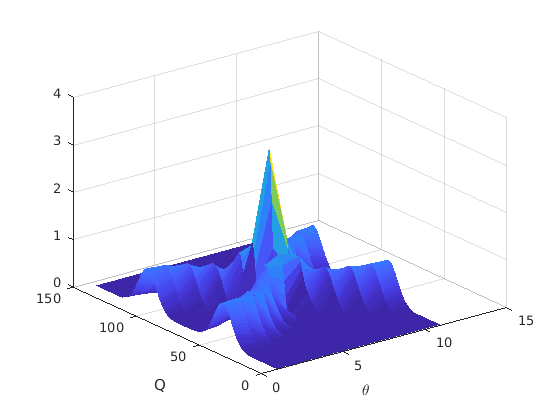}
        \caption{}
        \label{fig:radoncat10m}
    \end{subfigure}
     ~ 
    \begin{subfigure}{0.45\textwidth}
        \centering
        \includegraphics[width=\textwidth]{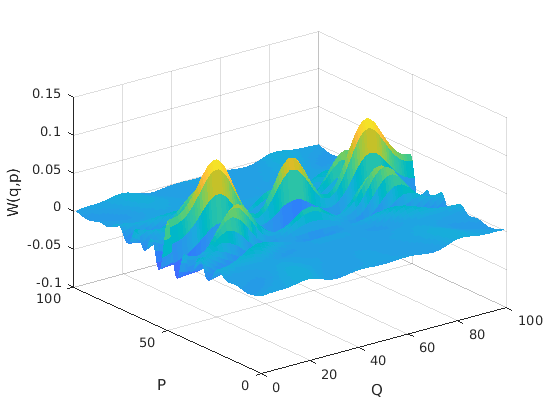}
        \caption{}
        \label{fig:reccat10m}
    \end{subfigure}
    ~ 
    \begin{subfigure}{0.45\textwidth}    
        \includegraphics[width=\textwidth]{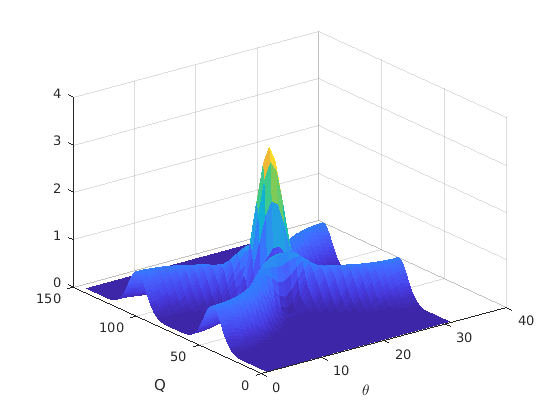}
        \caption{}    \label{fig:radoncat30m}
    \end{subfigure} 
     ~ 
    \begin{subfigure}{0.45\textwidth}
        \centering
        \includegraphics[width=\textwidth]{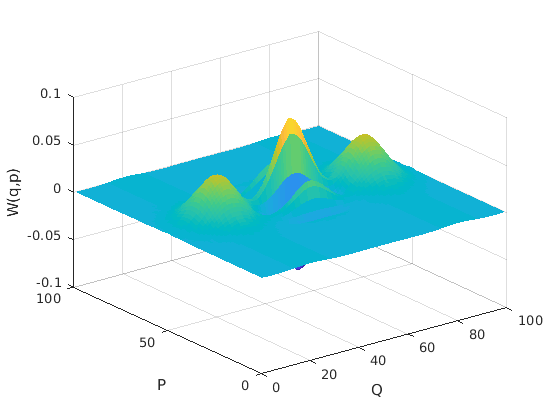}
        \caption{}
        \label{fig:reccat30m}
    \end{subfigure}
    \caption{Radom transforms with differents quantities of angles measured and the corresponding reconstructed Wigner function. The values range from 0 to 180 degress, with steps of 18 (a) and 6 (d) degress. Note on figure (b), the reconstruction has noise influence of the back-projection algorithm.}\label{fig:catradonrecs}
\end{figure}

We have seen that Wigner functions are useful to visualize the phase-space proprieties of quantum states: displays quadrature amplitudes, their fluctuations, and possible interferences. Now we present a simulated quantum tomography experiment to illustrate the whole procedure. To simulate the process of homodyne measure, since it is a Radon transform (\ref{eq:radontranstrool}) of the Wigner function, the MATLAB \texttt{radon} routine is sufficient if we treat the Wigner representation as any other two dimensional image. As a matter of fact, the matrix generated from the script can be mapped in a gray scale image, as  we can an see one example on figure \ref{fig:grayvac}. Since real measures has imperfections, to add some noise, a random matrix is summed at the result of the Radon transform. To illustrate it, I used the Schrödinger Cat state for of the figure \ref{fig:cat3} and reconstructed via \texttt{iradon} with the noise, displayed at figure \ref{fig:catradonrecs}.

\begin{figure}
    \centering
    \begin{subfigure}{0.3\textwidth}
    	\centering
    	\vspace{15pt}
        \includegraphics[width=\textwidth]{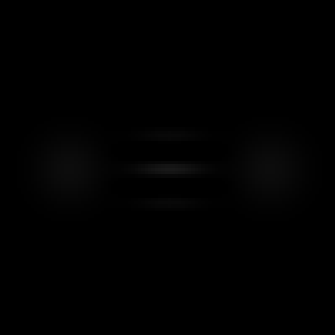}
        \vspace{7pt}
        \caption{Cat state as in gray scale}
        \label{fig:graycat}
    \end{subfigure}
    ~ 
    \begin{subfigure}{0.55\textwidth}
        \centering
        \includegraphics[width=\textwidth]{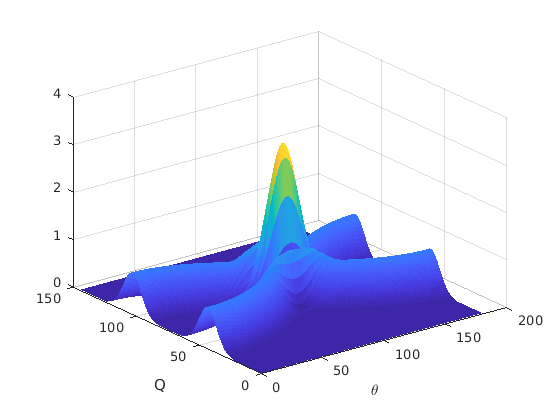}
        \caption{The Radon transform}
        \label{fig:radoncat}
    \end{subfigure}
     \caption{Radon transform of the Wigner function of the Cat state}\label{fig:catradon}
\end{figure}    
    
\begin{figure}   
    \begin{subfigure}{0.45\textwidth}
        \centering
        \includegraphics[width=\textwidth]{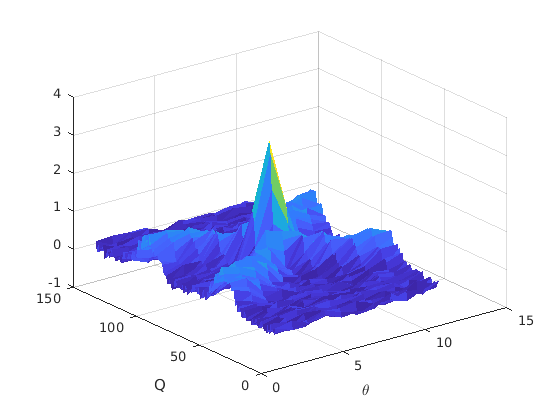}
        \caption{}
        \label{fig:noiseradoncat10m}
    \end{subfigure}
     ~ 
    \begin{subfigure}{0.45\textwidth}
        \centering
        \includegraphics[width=\textwidth]{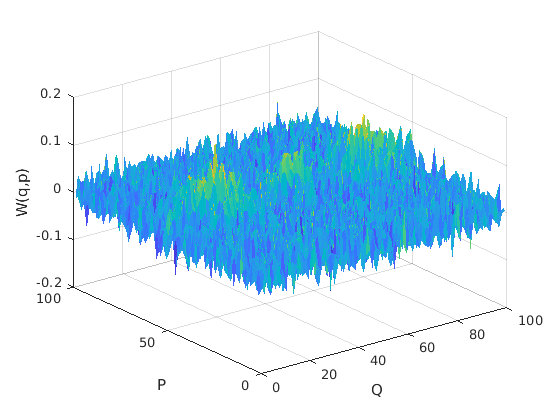}
        \caption{}
        \label{fig:noisereccat10m}
    \end{subfigure}
    ~ 
    \begin{subfigure}{0.45\textwidth}    
        \includegraphics[width=\textwidth]{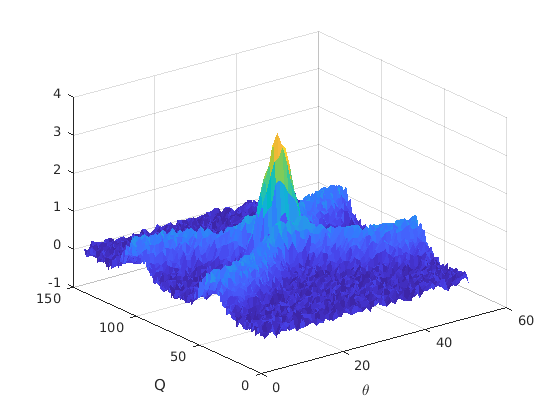}
        \caption{}
        \label{fig:noiseradoncat50m}
    \end{subfigure} 
     ~ 
    \begin{subfigure}{0.45\textwidth}
        \centering
        \includegraphics[width=\textwidth]{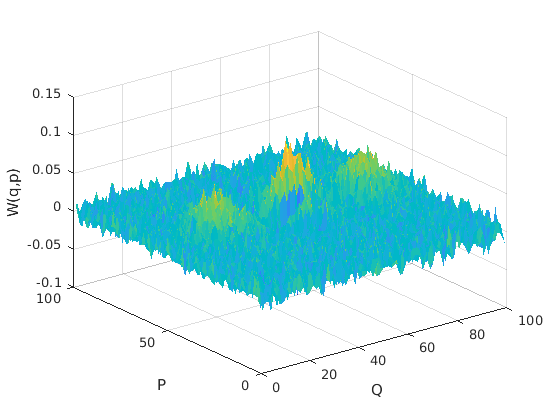}
        \caption{}
        \label{fig:noisereccat50m}
    \end{subfigure}
    \caption{Reconstructed Wigner function for the Cat state with noise}\label{fig:noisecatradonrecs}
\end{figure}

I have tested for different quantities of measurements, variating the number of angles $\theta$ and the noise of the quadratures measured values, as it is available on the figures \ref{fig:noisecatradonrecs}.

To perform the back-projections algorithm, since it has the same protocol form images, the \texttt{iradon} function on MATLAB suits an efficient cutoff on our case. The inverse Radon routine is already known on the fields of image treatment, for instance on medical tomography\footnote{The mathematics of tomography dates back 1917, with Johann Radon article \cite{radontransform}. It was latter finaly used in the early 1970's, given the Nobel prize in 1979 to  Cormarck and Hounsfield "for the development of computer assisted tomography".} \cite{hermanimage} and pattern recognition \cite{Illingworth1988}.

\subsection{Density operator in Fock Basis}

\begin{figure}
    \centering
    \begin{subfigure}{\textwidth}
    	\centering
    	\vspace{15pt}
        \includegraphics[width=\textwidth]{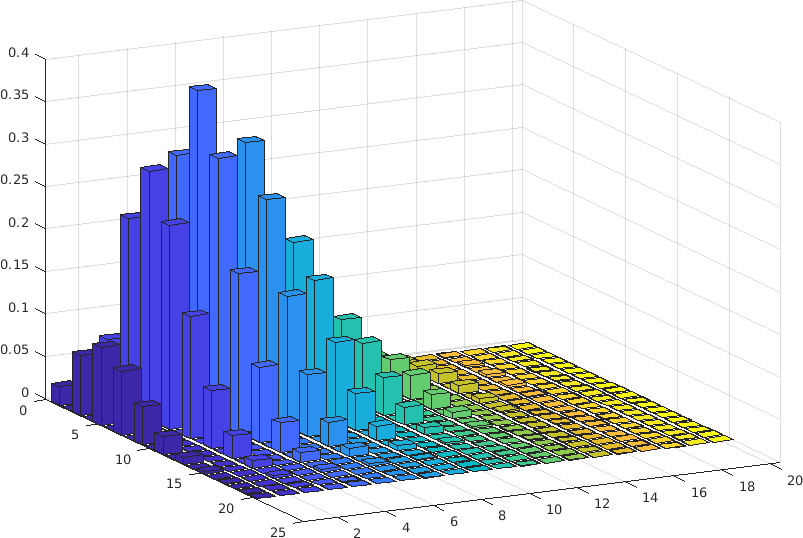}
        \vspace{7pt}
        \caption{Density matrix expressed in the Fock state basis}
        \label{fig:rhocat3}
    \end{subfigure}
    ~ 
    \begin{subfigure}{\textwidth}
        \centering
        \includegraphics[width=\textwidth]{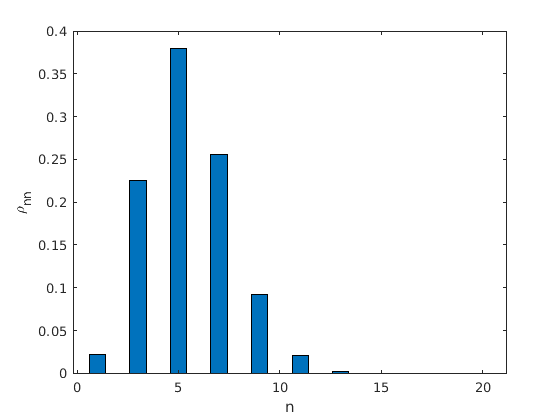}
        \caption{Diagonal elements}
        \label{fig:diagrhocat}
    \end{subfigure}
     \caption{Cat State matrix elements, $\bra{m}\rho\ket{n}$. Note that only when both $m$ and $n$ are even is the matrix element non-zero, because of the destructive interference between the odd Fock components of the two coherent states making up the Schrödinger cat.}\label{fig:fockcat}
\end{figure}

Using the relation (\ref{eq:matrixind}), I have create a program to make a list of Fock states on space representation functions and then, utilizing the same list, build projectors and their corresponding Wigner functions. Each projector is a $100\times100$ matrix, as the standard of this dissertation. With those tools, one can map the operator $\rho$ on a truncated matrix on Fock basis, as we can see on figure \ref{fig:fockcat}. We can map the Wigner state numerically using summation algorithm, here I used the MATLAB function for trapezoidal numerical integration \texttt{trapz}, for the best fit.

Since I have been mostly testing with coherent states, they are used to archive the best truncation for the density matrix. Expanding the coherent states as infinite sum of Fock states, as we see on (\ref{eq:coherentfock}), I decided that for $\alpha/n!$ must be sufficiently small, around $10^{-8}$. We tested the efficient of the algorithms on ranges from 10, 20, 31 Fock state basis, with the last one taking around 8h to complete the  projectors generation. Of course, the bigger is the basis used, better is the fit, but to produce it takes long time, to be more specific, $n^2$, since we need to combine all the $\ket{n}\bra{m}$ to correctly project the state.

Quantum state reconstruction can never be perfect, due to statistical and systematic uncertainties in the estimation of the measured statistical distributions. In both discrete and continuous variable domains, inverse linear transformation methods work well only when these uncertainties are negligible, \ie , in the limit of a very large
number of data and very precise measurements.

\begin{figure}
    \centering
    \begin{subfigure}{\textwidth}
    	\centering
    	\vspace{15pt}
        \includegraphics[width=\textwidth]{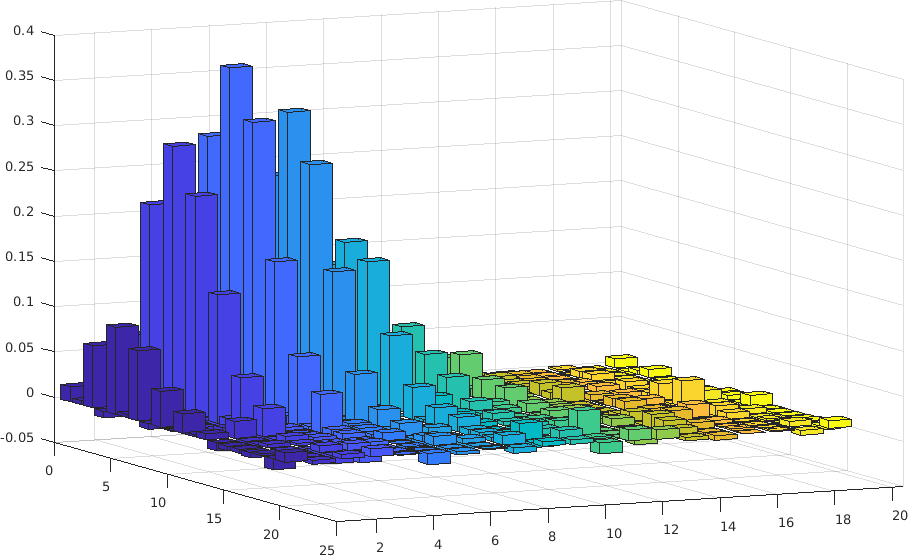}
        \vspace{7pt}
        \caption{}
        \label{fig:rhocat3noise}
    \end{subfigure}
    ~ 
    \begin{subfigure}{\textwidth}
        \centering
        \includegraphics[width=\textwidth]{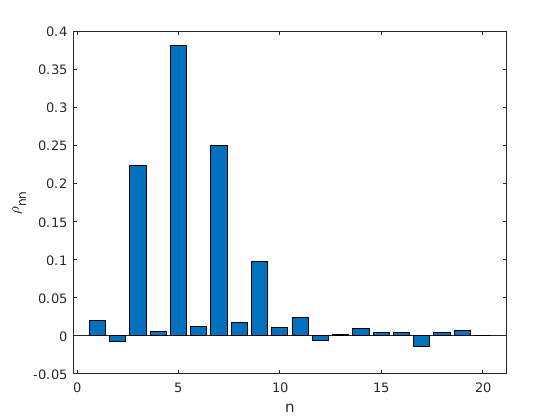}
        \caption{The presence of negative elements in the density matrix $\rho$ diagonal}
        \label{fig:diagrhocatnoise}
    \end{subfigure}
     \caption{Cat State reconstructed with noise density matrix expressed in the Fock state basis. Note the presence of negative elements in the diagonal of the matrix.}\label{fig:fockcatnoise}
\end{figure}

\subsection{Reconstructed States and Post-Processing}

As we can see, given the nature of the Wigner functions, the reconstruction does not lead necessarily to a state. It could be the lack of resolution, not sufficient measurements and also the result of noise. It is necessary to do a data post-processing to correctly estimate the state. The most popular way is to use \textit{Maximum Likelihood} (MaxLik) \cite{Lvovsky2003,Banaszek1998}. The basic idea of the method is to ask: ``what is the physically allowed state that most likely would have produced the observed distribution of quadratures''? This approach guarantees that the reconstructed state will be physically meaningful. The iterations will approach the global likelihood maximum. However, the coast for this kind of reconstruct is high, since it works as a global optimization of the state. We would like to try something different. One can suppose the correct state using the variational quantum tomography protocol \cite{Maciel2010}. The original technique was presented for discrete states. But once we write the state in a truncated Fock basis, it is the same processing.

So, the best we can do in this case is to use a semidefinite programs (SDPs). Since we can map our state on a good basis, for instance, the Fock basis, it give us the key to write a more reliable state. The SDPs are convex optimization problems which can be written as the minimization of a linear objective function, subject to semidefinite constraints in the form of linear matrix inequalities \cite{boyd2004convex}. From the definition by John Waltrus on his 2011 lecture \cite{WaltrousQI2011}:

\begin{defi}A semidifinite program is a triple $(\Phi,A,B)$, where
\begin{itemize}
\item[1]$\Phi \in T(\mathcal{X}, \mathcal{Y})$ is a Hermiticity-preserving linear map, and 
\item[2]$A \in \text{Herm}(\mathcal{X})$ and $B \in \text{Herm}(\mathcal{Y})$ are Hermitian operators,
\end{itemize}
for some choice of complex Euclidian spaces $\mathcal{X}$ and $\mathcal{Y}$.
\end{defi}

We associate with the triple $(\Phi,A,B)$ two optimization problems, called the \textit{primal} and \textit{dual} problems, as follow:

\begin{table}[H]
\begin{tabular}{llllllll}
\multicolumn{2}{c}{{\ul Primal Problem}}      &  &  &  &  & \multicolumn{2}{c}{{\ul Dual Problem}}         \\
maximize:   & $\langle A,X \rangle$           &  &  &  &  & minimize:   & $\langle B,Y \rangle$            \\
subject to: & $\Phi(X)=B$                     &  &  &  &  & subject to: & $\Phi^*(Y) \geq A$               \\
            & $X \in \text{Pos}(\mathcal{X})$ &  &  &  &  &             & $Y \in \text{Herm}(\mathcal{Y})$
\end{tabular}
\end{table}

Solving the dual problem, gives a lower bound on the primal problem. It is often the case that these values coincide — in which case, the SDP have the so-called strong duality property. Semidefinite programs have also another appealing feature: efficient algorithms are available for solving SPDs in polynomial time with arbitrary accuracy \cite{sturm1999using}. Than, given experimental reconstructed density matrix $\rho_{exp}$, the data (post-)processing can estimate efficiently the physical state. This can be done by means of the following very simple SDP: 

\begin{equation}\label{vqtinfhoro}
\begin{aligned}
\min_{\rho} \ \ \ & \quad \sum_{q,\theta} \abs{\trace{\rho \ketbra{q_\theta}{q_\theta}}-\mathrm{pr}(q,\theta)} \\
\text{s.t.} \ \ \  & \quad \rho \ \succeq 0 \, ; & \ \\
& \quad \trace{ \rho} = 1 \, ; & \  \\
\end{aligned}
\end{equation}

Note that we cannot use the SDP directly on the Wigner function, since the constrains would be far more difficult. The fact that the positive semidefinite state operator already has a series of constrains to be a actual quantum state, it bounds our problems.

\chapter{Aftermath: The Stories That Numbers Tell Us}

\epigraphfontsize{\small\itshape}
\epigraph{"To be honest is hard."}{--- \textup{Thiago Maciel}}

\section{Reconstructed state}

The Wigner function provide us with visual interpretation and statistic distribution of the state, but to confine the analysis on it would left us with the intrinsically error of measurement process plus the back-projection protocol. Furthermore, for more modes, the interesting visual features get lost and the error increases very fast.


Besides having infinite dimension, in this thesis, for a low photons state, our numerical investigations showed that a basis around 20 Fock states is enough to describe it in a accurate resolution. Using the Fock basis is a way to discretized an infinite basis state and avoids the complications of a position or momentum function.

We have gave preference on SDP estimation over the MaxLike because it is straightforward to implement and offers improvements over the inverse-linear-transform techniques such as inverse Radon. While maximum-likelihood wants to combine with maximum-entropy and Bayesian methods to improve the reconstruction \cite{Fuchs2004}, the SDP works with convexity: bounded problems and if the problem is feasible or not given a constrain. I think it is way simpler to write as an algorithm. I believe for a multi-mode state would be the fastest answer to correctly estimate the density matrix and, theoretically, it is also possible to correct for the detector inefficiencies \cite{Thiagoerror2017}.


\section{The Entanglement Resource}

Along the features of the density matrix, which tell us about the preparation of the state and probabilities, more information could be extracted on it. For instance, for a multipartite state, the cornerstone of quantum mechanics, the \textit{entanglement}. As stated by one of the founding fathers of quantum mechanics, Erwin Schrödinger on his paper from 1936 \cite{schrodinger1935}:

\begin{quote}
``I would not call (quantum entanglement) one but rather the characteristic trait of quantum mechanics, the one that enforces its entire departure from classical lines of thought.''
\end{quote}

But what is entanglement? What make it so special? Let us starting by definite it, since its extension to a multipartite scenario is simple, but with cumbersome notation, we will just present the definition for the bipartite case. Let $\mathcal{H}_A$ and $\mathcal{H}_B$ be a Hilbert spaces. Then we have:

\begin{defi}[Quantum Entanglement \cite{Werner1989}]\label{def:entanglement}
A quantum state $\rho_{AB}$ : $\mathcal{H}_A \otimes \mathcal{H}_B \rightarrow \mathcal{H}_A \otimes \mathcal{H}_B$ is separable if it can be written in the form 
\begin{equation}
\rho_{AB} = \sum_{\lambda} \pi (\lambda) \rho_{A}^{\lambda} \otimes \rho_{B}^{\lambda} \label{eq:separability}
\end{equation}
for some distribution $\pi$ : $\Lambda \rightarrow [0,1]$ and quantum states $\rho_{A}^{\lambda}$ : $\mathcal{H}_{A} \rightarrow \mathcal{H}_{A}$, $\rho_{B}^{\lambda}$ : $\mathcal{H}_{B} \rightarrow \mathcal{H}_{B}$.

Quantum state that are not separable are entangled.   
\end{defi}
In other words, a general quantum state of a two-party system is separable if its total density operator is a mixture, a convex sum of product states. 

In quantum information theory, entanglement is understood as a resource that can be used for protocols like, superdense coding \cite{bennett1992,Ban1999, Braunstein2003}, quantum teleportation \cite{Bennett1993,Milburn1999}, quantum cryptography \cite{Gisin2002,bennett1984quantum}, and possibly related to the exponential speed-up of quantum computation \cite{shor1997}.

The definition \ref{def:entanglement} is very easy but not practical. Decomposing the states into tensor products, as we can see at Eq. (\ref{eq:separability}), in order to show that a state is separable, is a very difficult and potentially lengthy task especially for high dimensional systems \cite{Eisert2003}.

Although, we can analyse the separability in a different and more efficient way using the theory of positive but not completely positive maps. For instance, considering the transposition applications, it maps a positive operator into a positive operator as well. And for a separable state, it is also valid for applying it to single subsystem, let us say subsystem $B$, once we get:
\begin{equation}
\rho^{T_B} = \sum_{i}p_i\left(\rho_i^A \otimes (\rho_i^B)^T \right),
\end{equation}
which is again a valid state. This operation is called \emph{partial transposition}. However, when we apply it to an inseparable state, then there is no certification that the result is again a positive operator, which means a physical state \cite{Eisert2003}. This is one of the entanglement criterion to be explored, the Peres-Horodecki criterion of positivity under partial transpose \cite{Peres1996,HORODECKI19961}\footnote{This criterion was introduced by Peres \cite{Peres1996} and shown to be necessary and sufficient for two qubits systems and for one quibit and one qutrit by the family Horodecki \cite{HORODECKI19961}.}. It says that if $\rho^{T_B}$ has negative eigenvalues, than $\rho$ is entangled. Although this criterion is capable of characterize a huge number of entangled states, it is only necessary for separability. It means that if it is PPT, we are not sure if is entangled or not, since the state can still be entangled, as the case of bound entanglement \cite{Horodecki1998}.

Despite important milestones in quantum information theory have been derived and expressed in terms of qubits or discrete variables, the notion of quantum entanglement itself came to light in a continuous-variable setting. For example, the two-mode squeezed state provides the physical realization \cite{eprreal} of Einsten-Podolsky-Rosen state \cite{EPRRRRRR} and is used as the prototype of a continuous variable for many quantum information protocols \cite{Braunstein2004}.

Let us discuss a little bit about entanglement on Gaussian States, that are already explored and used for many protocols.




\subsection{Entanglement on Gaussian States}
Many separability criteria has been proposal on last years, including for continuous variable systems \cite{Simon2000}. For Gaussian states, it can be formulate understanding how partial transpositions acts on the level of covariance matrices. Which leads for the following result:

\begin{teo}[Werner, 2001]\label{teo:sepgauss}
\cite{Werner2001} Be $\gamma$ the covariance matrix of separable Gaussian state. Then, the are covariance matrices $\gamma_A$ and $\gamma_B$ so that
\begin{equation}
\gamma \leq \begin{pmatrix}
\gamma_A & 0 \\
0 & \gamma_B
\end{pmatrix}.
\end{equation}
Conversely, if this condition is satisfied, the Gaussian state with covariance $\gamma$ is separable.
\end{teo}

Generalization of this result for many parts systems can be found in \cite{Giedke2001}. The importance of the theorem \ref{teo:sepgauss} is that it constitute a SDP, which leaves us with an numerical operational method for characterization of m-entanglement on many parts Gaussian states\footnote{The generalization of the theorem \ref{teo:sepgauss} for may parts systems and m-entanglement is also a SDP}. Another operational criterion for entanglement on two parts system is based in a certain non linear map application on the covariance matrix was introduced in \cite{Giedke2001a}.

There is a relatively simple manner to realize partial transposition in Gaussian States. One needs simply to revert the canonical variable moments belonging to the first part, while that positions and the other moments are left intact. In terms of the covariance matrix $\gamma$, that means to multiply $\gamma_{\alpha \beta}$ by $-1$ whenever $\alpha$ or $\beta$ correspond to the first part moments. Using the theorem \ref{teo:sepgauss}, it is shown on \cite{Werner2001} that, for Gaussian states where one mode belongs to one part and $N$ modes belongs to the other, the non positivity of the partial transposition is a necessary and sufficient for the entanglement existence.

\section{Outlooks and Conclusion}

Quantum physics of light has been developing along two parallel avenues: ``discrete-variable'' and ``continuous-variable'' quantum optics. The continuous-variable community dealt primarily with the wave aspect of the electromagnetic field, studying quantum field noise, squeezing and quadratures measuring. Homodyne detection was the primary tool for field characterization. The discrete-variable side of quantum optics concentrated on the particle aspect of light: single photons, dual-rail qubits, and polarization measuring.

The division of quantum optics is thus caused not by fundamental but by pragmatic reasons. The difference between these two domains boils down to the choice of the basis in which states of an optical oscillator are represented: either quadrature (position or momentum) or energy eigenstates.

Novel results in the discrete-variable domain, such as demonstration of entanglement, quantum tomography, quantum teleportation, etc., were frequently followed by their continuous-variable analogs and vice versa.

While measuring superoperators associated with a certain quantum process has been investigated theoretically \cite{ChuangNielsen1997} and experimentally \cite{Altepeter2003} for discrete variables for quite some time, the progress in the continuous-variable domain are still slow paced. This seems to be an important open problem, whose solution holds a promise to provide much more complete data on the quantum processes than current methods. Another open problem on continuous variable systems is about how to build an entanglement witness. 

In our numerical investigation, we saw that back-projection algorithms presents a series of problems on the cost and the error associated with the cutoff choice. As you increase the number of modes, the cutoff frequencies interferes with each other. Moreover, there is also the measurement errors, which also are include in a real state tomography. I believe that a good improvement for reconstruction techniques would be in fact using the data post-processing. Maximum likelihood is the most popular way to do it, however, not the last word in quantum state tomography algorithms. The MaxLike approach can have an enormous cost since it is a global optimization.

A new possibility is to explore the variational quantum tomography (VQT) protocol \cite{Maciel2010} on continuous variable state, which is already been used for reconstruction unknown quantum states out of incomplete and noisy information discrete low dimensional states \footnote{The method is a linear convex optimization problem, therefore with a unique minimum, which can be efficiently solved with semidefinite programs.}. Since cutoff means also you need to discretized the algorithm, reconstructing using a Fock state basis is not a silly choice, but powerful, since we now deal with error of measurement. The SDP algorithms are proven to be efficient. We can write many modes states with no complication using Fock basis. We can match the useful to the pleasant.

It is also interesting if with think also think about quantum process tomography, \ie, to use a topographically approach and to find out how the process can be described, using a known quantum states to probe a quantum process. Another feature to investigate is the entanglement detection, \ie, to build an efficient entanglement witness. I believe that using the VQT protocol associated with the knowledge gathered thought this dissertation, maybe we can archive it.





 
%
%
%
%
%
\bibliography{ludmila-tese}
\bibliographystyle{ieeetr}
\end{document}